\documentclass[twocolumn,traditabstract,longauth]{aa}
\usepackage{amssymb}
\usepackage{mathptmx}
\usepackage{natbib}
\bibpunct{(}{)}{;}{a}{}{,}
\usepackage[]{graphicx}
\usepackage{lscape}
\usepackage{txfonts}
\usepackage{verbatim}
\usepackage{url}
\defcitealias{suyu20}{Su20}
\defcitealias{schuldt20}{Sc20}

\begin{document}

\title{HOLISMOKES -- II. Identifying galaxy-scale strong gravitational lenses in Pan-STARRS using convolutional neural networks
  \thanks{Table 1 is only available in electronic form at the CDS via anonymous ftp to cdsarc.u-strasbg.fr (130.79.128.5)
or via {\tt http://cdsweb.u-strasbg.fr/cgi-bin/qcat?J/A+A/}.}}

\author{R. Ca\~nameras\inst{1}, S. Schuldt\inst{1,2}, S. H. Suyu\inst{1,2,3}, S. Taubenberger\inst{1}, T. Meinhardt\inst{4}, L. Leal-Taix\'e\inst{4}, C. Lemon\inst{5}, K. Rojas\inst{5}, E. Savary\inst{5}}

\institute{
Max-Planck-Institut f\"ur Astrophysik, Karl-Schwarzschild-Str. 1, 85748 Garching, Germany \\
{\tt e-mail: rcanameras@mpa-garching.mpg.de}
\and
Physik Department, Technische Universit\"at M\"unchen, James-Franck Str. 1, 85741 Garching, Germany
\and
Institute of Astronomy and Astrophysics, Academia Sinica, 11F of ASMAB, No.1, Section 4, Roosevelt Road, Taipei 10617,
Taiwan
\and
Technical University of Munich, Department of Informatics, Boltzmann-Str. 3, 85748 Garching, Germany
\and
Institute of Physics, Laboratory of Astrophysics, Ecole Polytechnique F\'ed\'erale de Lausanne (EPFL), Observatoire de Sauverny, 1290 Versoix, Switzerland
}

\titlerunning{HOLISMOKES -- II. Pan-STARRS lens search}

\authorrunning{R. Ca\~nameras et al.} \date{Received / Accepted}

\abstract{We present a systematic search for wide-separation (with Einstein radius $\theta_{\rm E} \gtrsim 1.5$\arcsec), galaxy-scale
  strong lenses in the 30\,000~deg$^2$ of the Pan-STARRS 3$\pi$ survey on the Northern sky. With long time delays of a few days to
  weeks, these types of systems are particularly well-suited for catching strongly lensed supernovae with spatially-resolved multiple
  images and offer new insights on early-phase supernova spectroscopy and cosmography. We produced a set of realistic simulations by
  painting lensed COSMOS sources on Pan-STARRS image cutouts of lens luminous red galaxies (LRGs) with redshift and velocity dispersion
  known from the sloan digital sky survey (SDSS). First, we computed the photometry of mock lenses in $gri$ bands and applied a simple
  catalog-level neural network to identify a sample of 1\,050\,207 galaxies with similar colors and magnitudes as the mocks. Second, we
  trained a convolutional neural network (CNN) on Pan-STARRS $gri$ image cutouts to classify this sample and obtain sets of 105\,760
  and 12\,382 lens candidates with scores of $p_{\rm CNN}>0.5$ and $>0.9$, respectively. Extensive tests showed that CNN performances
  rely heavily on the design of lens simulations and the choice of negative examples for training, but little on the network
  architecture. The CNN correctly classified 14 out of 16 test lenses, which are previously confirmed lens systems above the detection
  limit of Pan-STARRS. Finally, we visually inspected all galaxies with $p_{\rm CNN}>0.9$ to assemble a final set of 330 high-quality
  newly-discovered lens candidates while recovering 23 published systems. For a subset, SDSS spectroscopy on the lens central
  regions proves that our method correctly identifies lens LRGs at $z \sim 0.1$--0.7. Five spectra also show robust signatures of
  high-redshift background sources, and Pan-STARRS imaging confirms one of them as a quadruply-imaged red source at $z_{\rm s}=1.185$,
  which is likely a recently quenched galaxy strongly lensed by a foreground LRG at $z_{\rm d}=0.3155$. In the future, high-resolution
  imaging and spectroscopic follow-up will be required to validate Pan-STARRS lens candidates and derive strong lensing models.
  We also expect that the efficient and automated two-step classification method presented in this paper will be applicable to
  the $\sim$4~mag deeper $gri$ stacks from the Rubin Observatory Legacy Survey of Space and Time (LSST) with minor adjustments.}


\maketitle

\section{Introduction}
\label{sec:intro}

Strongly lensed systems with time-variable sources provide competitive probes of the Hubble constant $H_{\rm 0}$, which are independent
of cosmic microwave background (CMB) observations \citep{planck18} and the local distance ladder \citep{riess19,freedman19,freedman20},
and allow one to assess the significance of the current $H_{\rm 0}$ tension. The COSmological MOnitoring of GRAvitational Lenses
(COSMOGRAIL) and $H_{\rm 0}$ Lenses in COSMOGRAIL's Wellspring (H0LiCOW) projects \citep[e.g.,][]{suyu17,courbin18} have recently
established the capacity of combining time-delay measurements and robust strong lensing models to constrain $H_{\rm 0}$ and measured
$H_{\rm 0} = 73.3 ^{+1.7}_{-1.8}$~km~s$^{-1}$~Mpc$^{-1}$ in flat Lambda cold dark matter ($\Lambda$CDM) cosmology using six lensed quasars
\citep{wong20}. The seventh lens has been analyzed by the STRong lensing Insights into the Dark Energy Survey (STRIDES) collaboration
\citep{shajib20,buckleygeer20}, and a detailed study of systematic effects is presented by \citet{millon20} as part of the Time-Delay
COSMOgraphy (TDCOSMO) organization. Moreover, the first two strongly lensed supernovae (SNe) with spatially-resolved multiple images
have been detected in recent years; one core-collapse SN was found behind the strong lensing cluster MACS\,J1149.5+222.3
\citep[SN Refsdal,][]{kelly15}, and one type Ia SN was found behind an isolated lens galaxy \citep[iPTF16geu,][]{goobar17}. These
findings open new perspectives on future $H_{\rm 0}$ measurements with lensed SNe. These types of systems are indeed well-suited for
time-delay measurements given the smooth, nonerratic SNe light curves which require shorter high-cadence monitoring than lensed quasars,
and the possibility of reducing microlensing effects by focusing on the early, achromatic expansion phase a few weeks after explosion,
and by using color light curves \citep[][Huber et al., 2020, in prep.;]{suyu20,huber19,bonvin19,goldstein18}. For lensed type Ia SNe,
the standardizable intrinsic peak luminosity of the source is also a valuable input for breaking the mass-sheet degeneracy in lens
mass models \citep{falco85}. Constraints on $H_{\rm 0}$ have already been derived with SN Refsdal \citep{grillo18,grillo20} and they
illustrate the great potential of such measurements, in particular for galaxy-scale strong lens systems that have simpler lens mass
distributions than galaxy clusters. Lensed SNe with adequate image separations providing time delays of a few days to weeks are
particularly promising and are relatively less sensitive to microlensing effects \citep{suyu20,huber19}.

In addition to precise measurements of the Hubble constant, strongly lensed SNe are promising as they allow for early-phase SN
studies. Multiply-imaged SNe detected from the first image can be combined with strong lensing models to predict the time delays
and future SN reappearance, as was done for SN Refsdal, in order to trigger follow-up observations within a few days of explosion.
This is currently not feasible for unlensed SNe beyond the local Universe due to their late discovery near peak luminosity. Such
early-phase studies are particularly valuable when tackling the progenitor problem of type Ia SNe and disentangling the
single-degenerate \citep[][]{whelan73}, double-degenerate \citep[][]{tutukov81}, and additional scenarios that have been extensively
debated over the last decades. For core-collapse SNe, these observations are important in order to characterize the progenitor
properties and compare them with current stellar evolution models. Early-phase spectroscopy of type II SNe would yield novel
constraints on the mass-loss history just before explosion.

We recently initiated the Highly Optimized Lensing Investigations of Supernovae, Microlensing Objects, and Kinematics of Ellipticals
and Spirals \citep[HOLISMOKES,][]{suyu20} program to address these fundamental questions on stellar physics and cosmology. The
number of strongly lensed SNe is expected to grow over the next few years, thanks to the on-going Zwicky Transient Facility
\citep[ZTF,][]{masci19} high-cadence survey on the Northern Hemisphere and the forthcoming Rubin Observatory Legacy Survey of Space
and Time \cite[LSST,][]{ivezic19} on the South. \citet{oguri10} predict 45 strongly lensed type Ia SNe over the ten years of LSST,
which corresponds to a few events for the shallower ZTF survey. This assumes a selection from spatially-resolved multiple images
targeting the most useful wide-separation systems. Using complementary selection techniques solely based on magnification of SN Ia
light curves, \citet{goldstein17} predict ten to 20 times more lensed SN Ia albeit mostly with small image separations \citep[see
  also][]{wojtak19}. Importantly, new lensed SNe candidates have to be selected early enough to start the follow-up sequence in a
timely manner. One way is to extend the numerous, successful searches of galaxy-scale strong lenses that were traditionally conducted
on surveys with optimal image quality \citep[e.g.,][]{more16,sonnenfeld18}, to surveys with largest sky coverage, in order to quickly
identify transients matching the position of background lensed sources. Ultimately, these lens finding pipelines will be directly
applicable to the deep LSST stacks which are expected to yield approximately a hundred thousand new systems \citep{collett15}.

Galaxy-scale strong gravitational lenses without time-variable sources also provide valuable insights into the lens total mass
distributions, including the inner dark-matter fractions \citep[e.g.,][]{gavazzi07,grillo09,sonnenfeld15,schuldt19}, the slopes
of the total and dark-matter mass density profiles \citep[e.g.,][]{treu02,koopmans09,barnabe11,shu15}, and the spatial extent of
dark-matter halos \citep[e.g.,][]{halkola07,suyu10}. Such systems play a crucial role in characterizing the lens stellar initial
mass function (IMF), a major ingredient for stellar mass estimates, as a function of galaxy physical properties
\citep[e.g.,][]{canameras17a,barnabe13,sonnenfeld19}, and they are well-suited to search for dark-matter
substructures \citep[e.g.,][]{vegetti12,hezaveh16,ritondale19}. Moreover, high magnification factors provide unique diagnostics
on the local interstellar medium physical conditions in lensed high-redshift galaxies and on the local feedback mechanisms
driving their evolution \citep[e.g.,][]{danielson11,canameras17b,cava18}.

Strong lensing events are rare, about one in 1000 for high-resolution space-based imaging \citep[e.g.,][]{marshall09} and down
to about one in 10$^5$ for seeing-limited ground-based data \citep[e.g.,][]{jacobs19b}. Thus, their identification requires
dedicated and automated methods. For instance, arc-finder algorithms \citep[e.g.,][]{gavazzi14,sonnenfeld18} and citizen-science
classification projects \citep[{\sc Space Warps},][]{marshall16} have been developed over the last decade. In particular,
convolutional neural networks (CNNs) are supervised machine-learning algorithms optimized to image analysis \citep{lecun98}
that have proven to outperform other classification techniques and that require little preprocessing. They are very efficient
to peer into large imaging data sets and have been increasingly used in the field of astronomy over the last five years. These
studies have established the ability of CNNs in recognizing galaxy morphologies \citep[][]{dieleman15}, including the key
features of strong gravitational lenses \citep[][]{metcalf19}. Several CNN searches for new strong lens candidates have focused
on ground-based imaging data, from the CFHTLS \citep{jacobs17}, KiDS DR3 \citep{petrillo17} and DR4 \citep{petrillo19a,li20},
DES Year 3 \citep{jacobs19a,jacobs19b}, or the DESI DECam Legacy survey \citep{huang19}. Efficient classification pipelines using
deep neural networks have also been developed and tested on simulated Euclid and LSST images to prepare for these forthcoming
surveys which will tremendously increase the number of detectable strong lensing systems
\citep[][]{lanusse18,schaefer18,davies19,cheng19,avestruz19}.

In the meantime, no systematic searches of galaxy-galaxy strong lenses have so far taken advantage of the Pan-STARRS imaging
covering the entire Northern sky. With this survey, \citet{berghea17} identified a strongly lensed QSO forming a quadruple
system by cross-matching the position of a variable AGN selected in the mid-infrared. More recently, \citet{rusu19} performed a
more systematic search of lensed QSOs by applying color and magnitude cuts and visually inspecting Pan-STARRS image cutouts of
AGN candidates from the Wide-field Infrared Survey Explorer \citep{secrest15}. In this paper, we perform a comprehensive search
for galaxy-scale strong lensing systems with luminous red galaxies (LRGs) as deflectors and typical high-redshift galaxies as
background sources, using the extended footprint of nearly 30\,000~deg$^2$ of the Pan-STARRS 3$\pi$ survey on the Northern sky.
The automated pipeline, based on a catalog-level preselection of galaxies and a convolutional neural network, results in a
ranked list of candidates which we further inspect visually to select those with higher confidence for spectroscopic follow-up.

The outline of the paper is as follows. In Sect.~\ref{sec:ps1} and \ref{sec:method}, we give a short overview of the Pan-STARRS
surveys and the overall search methodology and, in Sect.~\ref{sec:simu}, we describe the simulation of strong lenses. In
Sect.~\ref{sec:cnn}, we present the networks and training processes, and we extensively test the CNN performance. In
Sect.~\ref{sec:results} we finally apply the CNN to preselected Pan-STARRS image cutouts, provide the list of strong lens
candidates from visual inspection, and characterize their overall properties. Our main conclusions appear in
Sect.~\ref{sec:conclu}.  Throughout this work, we adopt the flat concordant $\Lambda$CDM cosmology with $\Omega_{\rm M}=0.308$,
and $\Omega_\Lambda=1-\Omega_{\rm M}$ \citep{planck16}, and with $H_0=72$~km~s$^{-1}$~Mpc$^{-1}$ \citep{bonvin17}.

\begin{figure*}
\centering
\includegraphics[width=.48\textwidth]{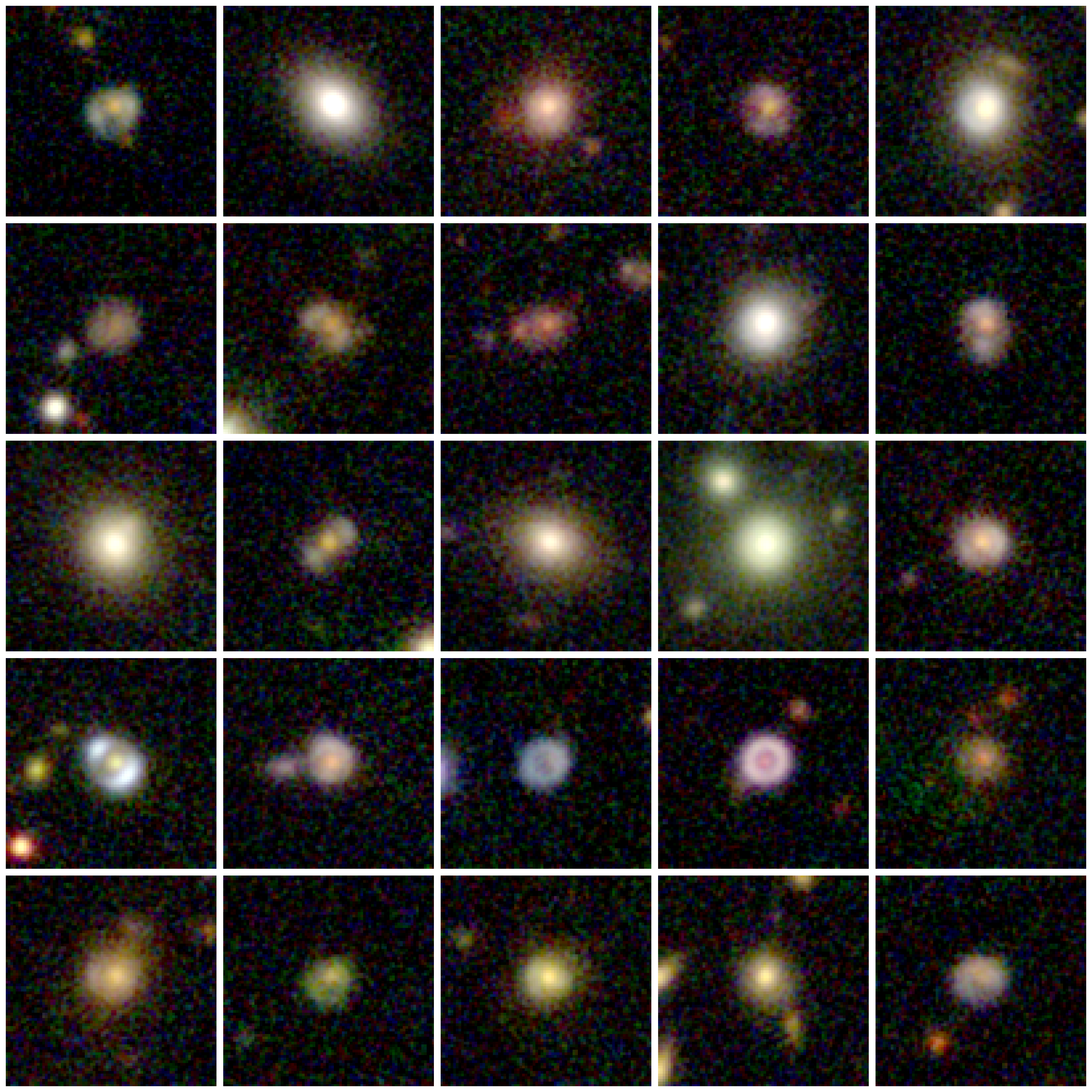}
\hspace{3mm} \includegraphics[width=.48\textwidth]{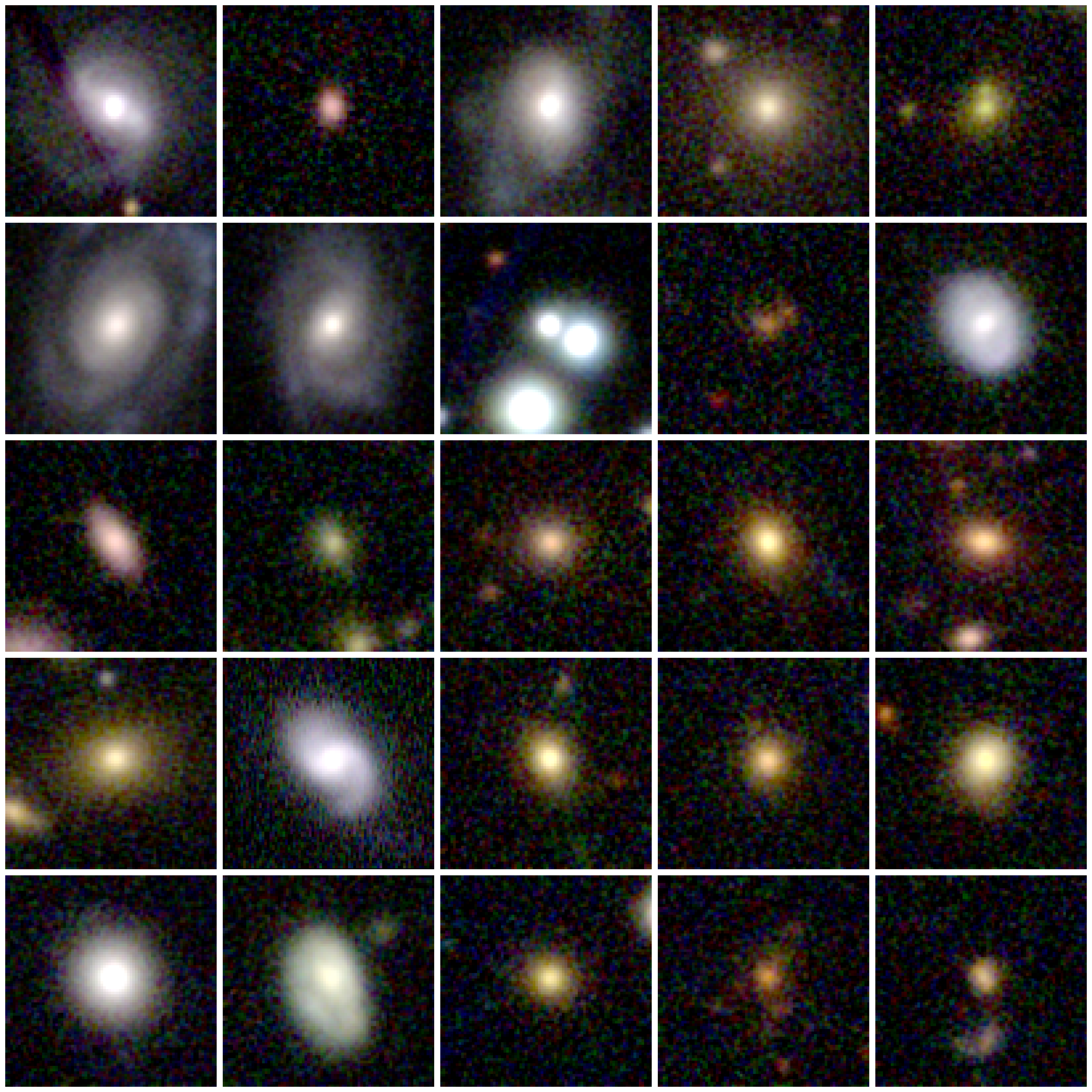}
\caption{
  {\it Left:} Examples of strong gravitational lens systems mocked up by painting COSMOS lensed sources on PS1 stack images, and
  used as positive examples for training. {\it Right:} PS1 postage stamps of a subset of the 90\,000 galaxies used as negative
  examples, including face-on spirals, massive LRGs, and field galaxies with similar $gri$ colors as the mocks. All postage stamps
  are 20\arcsec~$\times$~20\arcsec.}
\label{fig:simu}
\end{figure*}

\section{The Pan-STARRS1 survey}
\label{sec:ps1}

The Pan-STARRS1 (PS1) surveys were conducted with a 7~deg$^2$ field-of-view camera mounted on a 1.8~m telescope near the Haleakala
summit, Hawaii, in the five broadband $grizy$ filters \citep{chambers16} similar to those from the sloan digital sky survey (SDSS).
The camera has a pixel size of 0.258\arcsec~pix$^{-1}$ \citep{tonry09}. PS1 includes both the 3$\pi$ and Medium Deep surveys. The
former was completed in 2014 and made publicly available in DR1 and DR2. It covers 30\,000~deg$^2$ on the Northern sky down to
--30~deg in the five $grizy$ filters with a depth of 21--23~mag. The median seeing FWHM is 1.31\arcsec, 1.19\arcsec, and
1.11\arcsec\ in $g$, $r$, and $i$ bands, respectively, but reaches $>$1.60\arcsec, $>$1.45\arcsec, and $>$1.35\arcsec\ over 20\% of
the footprint \citep{chambers16}. The Medium Deep survey consists of ten fields covering a total of 70~deg$^2$, with multiple visits
in five filters optimized for transient detection. The few hundred exposures will eventually provide deep stacks with 5$\sigma$ point
source detection limits down to $i \sim 26.0$~mag and will be available in DR3.

We performed our search on the entire Northern sky with the 3$\pi$ survey that overlaps nicely with on-going optical time-domain
surveys on the Northern Hemisphere \citep[e.g., ZTF,][]{masci19} which provide a wealth of astronomical transients, including
strongly-lensed SNe. PS1 extends the SDSS to lower declinations and achieves higher depth. In particular, we applied our
pipeline to $gri$ stack images from DR1 which provide the optimal coaddition of individual exposures and have higher 5$\sigma$
point-source sensitivities than $z$ and $y$, probing down to 23.3, 23.2, and 23.1 mag in $g$, $r$, and $i$, respectively
\citep{chambers16}. These three bands conveniently span a wavelength range sensitive to young stellar populations in blue
star-forming galaxies at $z \gtrsim 1$ and to more dust-obscured or evolved early-type galaxies. Although the overall survey
strategy provided limited variations in depth and image quality over large scales, 3$\pi$ images have non uniform coverage
on small scales due to the stacking process. We found that limiting magnitudes vary by up to $\sim$0.5~mag on PS1 cutouts
over the extragalactic sky and accounted for this effect in our analysis.

\section{Overview of the lens-search method}
\label{sec:method}

In this paper, we aim at identifying galaxy-scale strong lensing systems on the extragalactic sky covered by PS1. We focused our
search on typical high-redshift galaxies strongly lensed by massive LRGs, which have a higher lensing cross-section \citep{turner84}
and smooth light profiles that help separate the foreground and background emissions. In particular, given the long standing
difficulty in distinguishing strong lensing features from arms of low redshift spirals, lenticular galaxies, tidal tails and
other contaminants with arc-like features \citep[e.g.,][]{huang19,jacobs19b}, restricting to lens LRGs increases our chance of
robustly identifying multiple lensed images with the $>1$\arcsec\ average PSF FWHM of PS1. 

Selecting these rare systems on the entire Northern sky requires an efficient analysis of the properties of the three billion sources
detected in the PS1 3$\pi$ survey image stacks. To circumvent memory limitations, a number of CNN searches in the literature
have focused on subsets of galaxies with LRG-like photometry using, for instance, the Baryon Oscillation Spectroscopic Survey
(BOSS) sample \citep{schlegel09,dawson13} or dedicated color and magnitude $gri$ cuts adapted from \citet{eisenstein01}. However,
this approach requires low contamination from lensed images to the lens multiband photometry and was essentially applied to deeper
surveys with better image quality (subarsec PSF FWHM in optical bands) than Pan-STARRS, such as the Hyper Suprime-Cam Subaru
Strategic Program \citep[HSC-SSP,][]{aihara18,sonnenfeld18} or the Kilo-Degree Survey with OmegaCAM on the VLT Survey Telescope
\citep[KiDS,][]{dejong13,petrillo19a}. Due to the lower image quality, applying such simple cuts on the photometry tabulated
in Pan-STARRS DR2 catalogs would exclude significant fractions of interesting systems with strongly lensed arcs blended with
the lens and altering its photometry. We therefore adopted a two-step approach: (1) a catalog-based neural network classification
of source photometry, (2) a CNN trained on $gri$ image cutouts.

In addition, we aim to find wide-separation lens systems because these configurations provide longer time delays of a few
days to weeks between multiple images, which is crucial to measure accurate, microlensing-free time delays for cosmology
\citep{huber19}. Recently, the extensive follow-up of the lensed Type Ia SN iPTF16geu at $z=0.4$ with
$\theta_{\rm E} \sim 0.3$\arcsec\ \citep{goobar17} illustrated the difficulty in reaching the time-delay precision required for
cosmography on small-separation systems \citep[$\Delta t < 2$~days,][]{dhawan19}, and demonstrated the impact of microlensing
\citep{more17,yahalomi17,bonvin19}. Focusing on wide separations is an effective search strategy for the Pan-STARRS survey with
limited angular resolution, and it will help to trigger timely imaging and spectroscopic follow-up. 

As further described in Sect.~\ref{sec:cnn}, CNNs capture image characteristics by learning the coefficients of convolutional
filters (kernels) of given width and height and creating a range of feature maps. They are invariant to translation and rotation.
During the learning phase, the CNNs rely on training sets with representative labeled images to minimize the difference between
predictions and ground truth. Classification algorithms require training sets of a few 10$^4$ to a million of labeled images
depending on the number of classes, image complexity and network depth. In contrast to the recent computer vision image recognition
challenges using deep CNNs \citep[e.g.,][]{russakovsky15}, relatively modest training sets of few 10$^5$ examples are sufficient
for our two-class problem applied to small, galaxy-scale image cutouts \citep[e.g.,][]{jacobs17,jacobs19b}. Nonetheless, the small
number and heterogeneous properties of spectroscopically-confirmed strong lenses (see the MasterLens database) make it necessary
to use simulated systems.

Generating realistic mocks that account for the complexity of PS1 stack images is a critical ingredient to reach optimal
classification performances \citep[e.g.,][]{lanusse18}. We constructed our mocks by painting lensed arcs on PS1 $gri$
images of LRGs with known redshift and velocity dispersion from SDSS spectroscopy. This approach captures the 3$\pi$ survey
properties, such as background artifacts, the presence of line-of-sight neighboring galaxies, and local variations of seeing
FWHM, exposure time and noise levels, while also accounting for variations in individual bands. In contrast to fully-simulated
images, using real cutouts also guarantees positive examples that best mimic the small scale background properties inherited from
the complex masking and stacking of individual PS1 exposures. As background sources, we used representative high-redshift galaxies
from the COSMOS field. High S/N image cutouts were taken from HSC-SSP and Hubble Space Telescope (HST) to simulate lens distortions
and magnifications, and we used $gri$ bands (similar filter set as PS1) to provide color information.

In Sect.~\ref{sec:simu}, we present the selection of lens and source galaxies and the pipeline to produce a set of mocks. Our
first catalog-level network, described in Sect.~\ref{ssec:catsearch}, was trained on the multiband photometry of mocks and
nonlens systems from the PS1 catalog, and assigned an output score, $p_{\rm cat}$, ranging between 0 and 1. A much lower fraction
of sources with $p_{\rm cat}>0.5$ were then classified with the CNN presented in Sect.~\ref{ssec:training}, resulting in the final
score, $p_{\rm CNN}$. We eventually examined visually all sources with highest scores to assign grades and collected a list of
high-confidence candidates for future validation.

\section{Simulating galaxy-scale strong lenses}
\label{sec:simu}

\subsection{Selection of lens galaxies}
\label{ssec:lens}

Realistic strong lensing simulations require knowledge on the lens mass distribution and redshift. We therefore
drew our sample of lens LRGs from the SDSS spectroscopic samples with reliable velocity dispersion measurements, to have a
proxy of the lens total mass. We used the SDSS large scale structure catalogs of galaxies and QSOs for cosmological studies,
including LOWZ and CMASS samples for BOSS (from SDSS DR12), and the higher-redshift LRG catalog for eBOSS
\citep[from SDSS DR14,][]{bautista18}. QSOs were excluded using SDSS {\tt class} flag. This resulted in a broad sample of
LRGs selected for their redder rest-frame colors using $gri$ color and magnitude cuts \citep[][]{eisenstein01}. The sample is
volume limited up to $z \sim 0.4$ (LOWZ), with additionally more luminous LRGs in the range $0.4 < z < 0.7$ (CMASS). eBOSS
LRGs lie at higher redshift \citep[$z_{\rm med} \sim 0.7$,][]{prakash16} due to a combination of optical and mid-infrared cuts
in SDSS and WISE bands.

We cleaned this spectroscopic catalog to keep LRGs with reliable velocity dispersions, using $v_{\rm disp} \leq 500$~km~s$^{-1}$
and $v_{\rm disp,err} \leq 100$~km~s$^{-1}$, and obtained 1\,192\,472 LRGs to build the mocks. We then cross-matched with the PS1
catalog to obtain their photometry, image depth and seeing FWHM in PS1.

\subsection{Selection of background sources}
\label{ssec:source}
      
The sample of galaxies used to mock up high redshift lensed sources was drawn from the COSMOS field to take advantage of the 
wealth of existing data including ultra-deep optical imaging, multiband photometry, spectroscopic follow-up, and morphological
classification. We selected galaxies with morphological information from {\sc Galaxy Zoo: HST} \citep{willett17} and within 
the COSMOS2015 photometric catalog \citep{laigle16} that also lists physical parameters from SED fitting. The former is a 
citizen science project that extends the original Galaxy Zoo \citep{lintott08,lintott11,willett13} with a thorough visual
classification of galaxies with ACS imaging from the Hubble legacy surveys \citep[see][for the COSMOS 
field]{scoville07,koekemoer07}. In particular for COSMOS, {\sc Galaxy Zoo: HST} relies on 3-color images obtained by 
combining the HST F814W mosaic with color gradients from ground-based imaging\footnote{Galaxy Zoo: CANDELS \citep{simmons17}
  performs similar classifications using deeper, multiband HST images but only over a subset of the COSMOS field
  \citep{grogin11,koekemoer11} and strongly restricts the sample size.}.

We cleaned the resulting catalog from sources identified as stars or artifacts (COSMOS2015 flag or visual identification),
and removed very extended galaxies with $R_{\rm eff} > 1.5$\arcsec, as well as galaxies contaminated by emission from
companions within 5\arcsec, and brighter by 1~mag in $r$ band \citep{laigle16}. The output sample included 52\,696 galaxies
for the strong lensing simulations. Redshifts were taken from public spectroscopic redshift catalogs drawn from surveys with
VLT/VIMOS \citep[zCOSMOS-bright,][]{lilly07}, VLT/FORS2 \citep{comparat15}, Subaru/FMOS \citep{silverman15}, VLT/VIMOS
\citep[VUDS,][]{lefevre15,tasca17}, Keck/DEIMOS \citep{hasinger18}, or the best photometric redshift estimate from
\citet{laigle16} for galaxies without $z_{\rm spec}$ available.

For the purpose of using this pipeline in future lensed SN searches, the properties of COSMOS sources were compared with
expectations for high-redshift SN hosts. Firstly, the cleaned catalog has a redshift distribution peaking at $z \sim 0.8$
with a tail extending to $z \gtrsim 1.5$, akin to the mock lensed SN catalog of \citet{oguri10}, both for LSST-like imaging
or for current, shallower surveys probing down to $R \sim 20$--21 \citep[e.g., ZTF,][]{masci19}. Secondly, the morphologies
and star formation activities can be put in context with properties of SN hosts constrained in the local Universe. Using
a compilation of $>$3000 SN and host properties from SDSS-DR8, \citet{hakobyan12} show that $\sim$13\% of SNe of all
types at $z \lesssim 0.1$ explode in galaxies with elliptical or lenticular morphologies, typical of early-type hosts. We
augmented the quiescent vs. star-forming classes of \citet{laigle16} based on redshift-dependent ${\rm NUV}-r/r-J$ cuts,
with Galaxy Zoo morphologies to conclude that our sample is strongly dominated by star-forming galaxies, with $\sim$15\%
classified as quiescent. This shows that our sample of sources broadly matches the expected properties of SN hosts.

\begin{figure*}
\centering
\includegraphics[height=.40\textwidth]{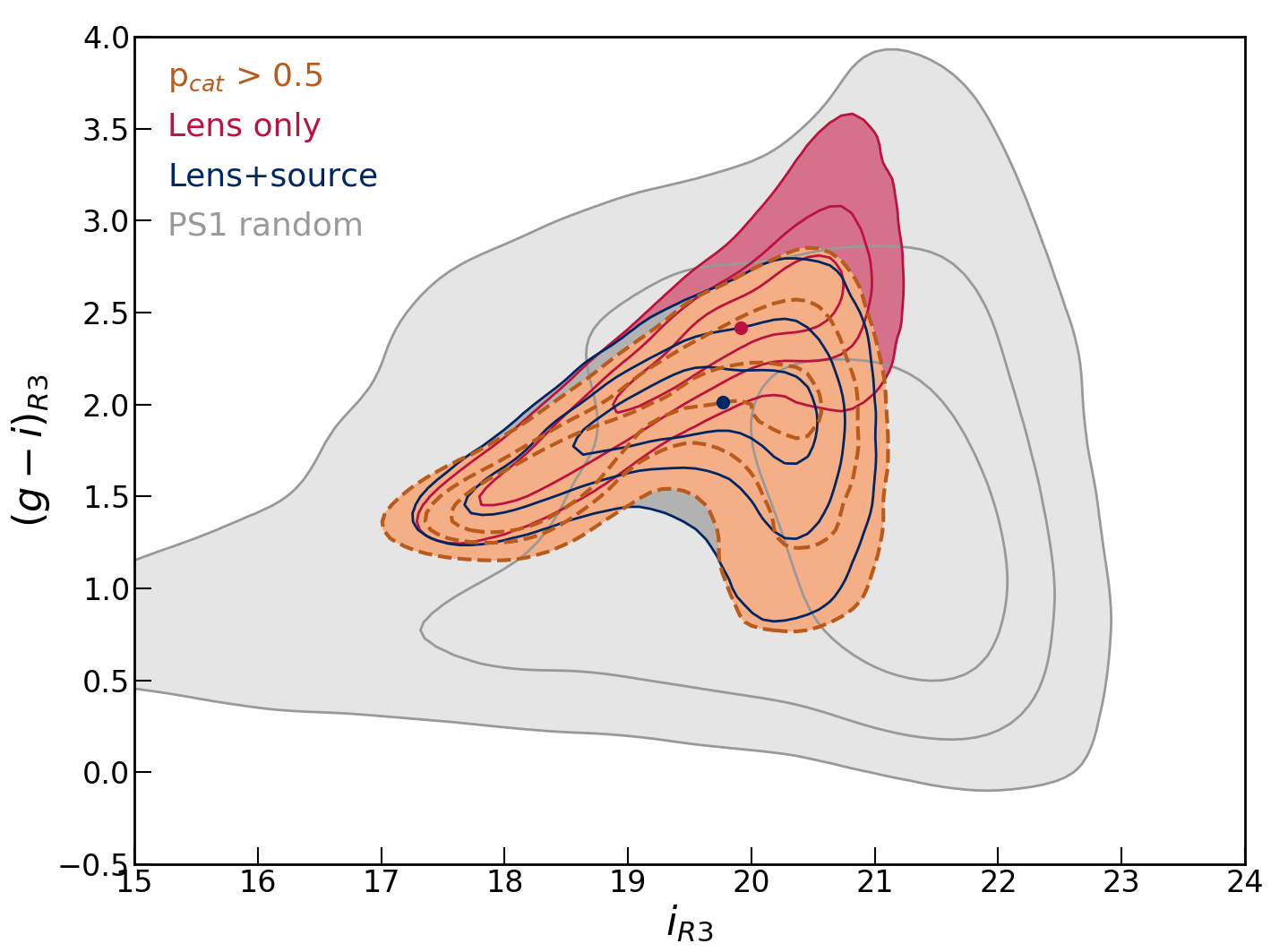}
\includegraphics[height=.40\textwidth]{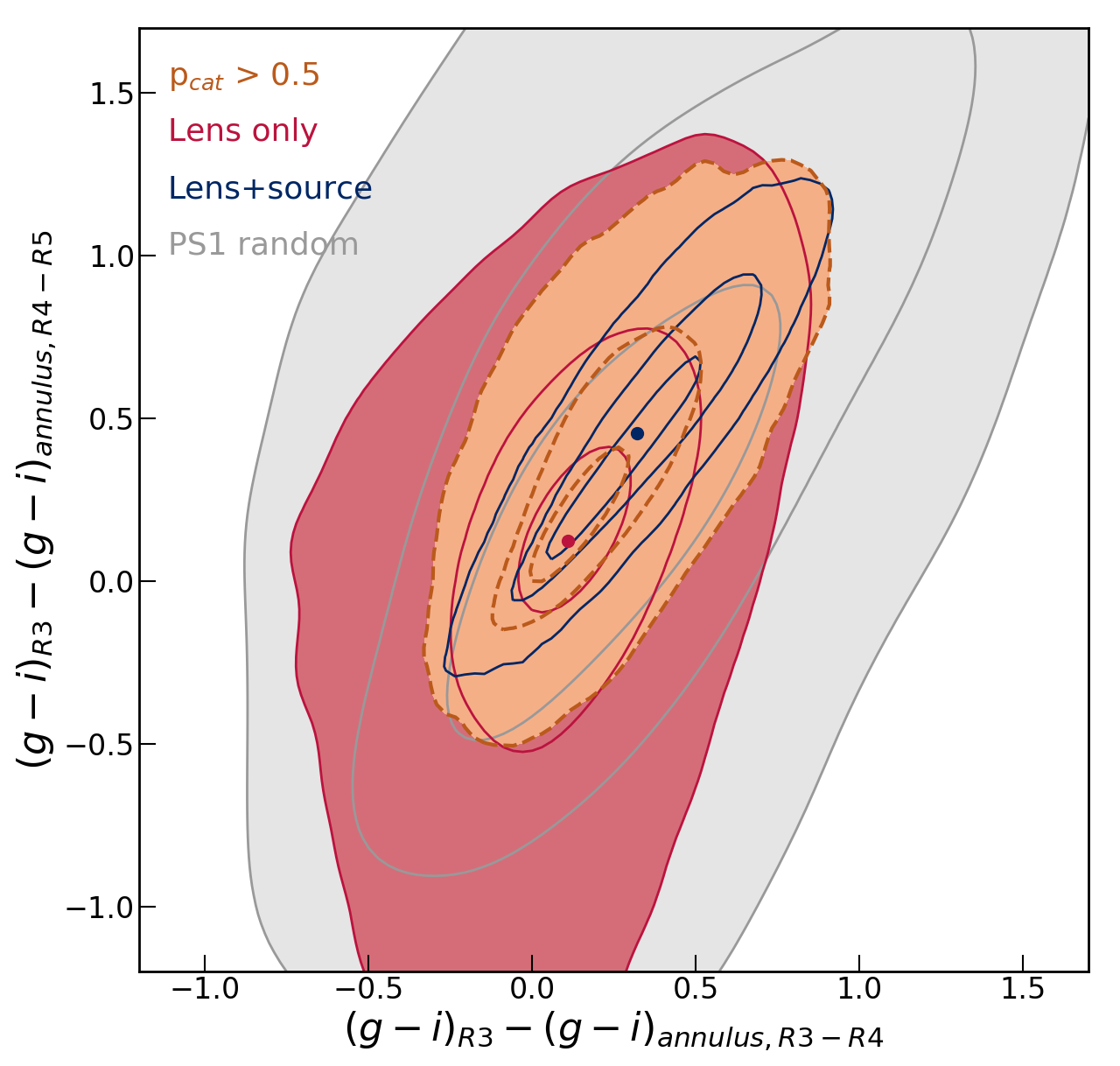}
\caption{
  Aperture magnitudes and colors of galaxies in the lens LRG catalog (red regions) and in the set of 90\,000 mock lens systems
  (blue regions). The red and blue dots show the median of distributions. {\it Left:} $(g-i)$ vs. $i$ diagram for the R3 aperture,
  a circular aperture of 1.04\arcsec\ radius. {\it Right:} Difference in $(g-i)$ color between the inner R3 aperture and concentric
  annuli between R3 and R4 (1.76\arcsec\ outer radius), and between R4 and R5 (3.00\arcsec\ outer radius). Gray regions mark the
  position of 100\,000 random sources with reliable $gri$ aperture photometry selected from the PS1 DR2 catalogs. Orange regions
  show galaxies with $p_{\rm cat}>0.5$ on the catalog-level network, which match the colors and magnitudes of mocks lens systems.
  Solid and dashed lines show the 0.5, 1, and 2$\sigma$ contours.}
\label{fig:catsearch}
\end{figure*}

\subsection{Downloading and processing image cutouts}
\label{ssec:mask}

For the lenses, PS1 $gri$ image cutouts of 20\arcsec~$\times$~20\arcsec\ were downloaded from the PS1 cutout
service\footnote{http://hla.stsci.edu/fitscutcgi\_interface.html}. We characterized the image depth in individual cutouts
with SExtractor \citep{bertin96} and verified that the observing strategy leads to nearly uniform depth. Although it depends 
on several observing factors, the depth is weakly correlated with the number of individual warp exposures used in a given 
stack. In particular, it rapidly drops by 0.2--0.3~mag for the 10\% of cutouts obtained by coadding less than 8, 10, and 12 
frames in $g$, $r$, and $i$ bands, respectively. A small fraction of $<$5\% of these PS1 images were discarded from the analysis.

Multiband images of COSMOS galaxies used as lensed sources were taken from the first data release of the HSC SSP \citep{aihara18}.
The HSC ultra-deep stacks are providing the deepest optical exposures with best image quality over the 2~deg$^2$ of COSMOS and
are well suited for the simulation pipeline. The 5$\sigma$ point-source sensitivities are 27.8, 27.7, 27.6~mag, in $g$, $r$, $i$,
respectively, and seeing conditions are excellent, with median values of 0.92\arcsec, 0.57\arcsec, and 0.63\arcsec\ in $g$, $r$,
and $i$ bands, and negligible variations over the COSMOS field \citep{tanaka17}.

We downloaded $gri$ cutouts of 10\arcsec~$\times$~10\arcsec, sufficient to enclose all emission from galaxies with
$R_{\rm eff} < 1.5$\arcsec. Fainter companions within a few arcsec were masked using segmentation maps created in $r$ band
with SExtractor \citep[using relatively few deblending subthresholds,][]{bertin96} to isolate the central galaxy of
interest. To overcome the limited spatial resolution of ground-based images, we combined these frames with the HST F814W
high-resolution images over the COSMOS field \citep[see][]{leauthaud07,scoville07,koekemoer07} to produce pseudo color
images following the steps described in \citet{griffith12}. First of all, F814W images were aligned and rescaled as if observed
in HSC $i$ band, and masked HSC frames were resampled with SWarp \citep{bertin02} to the HST scaling of 0.03\arcsec~pix$^{-1}$ 
using nearest-neighbor interpolation. Secondly, we multiplied each resampled frame by an illumination map, defined as F814W
divided by HSC $i$ band. This process preserves HSC source photometry, and results in $gri$ images with high-resolution light
profiles and color gradients with seeing-limited resolution.

\subsection{Strong lensing simulations}
\label{ssec:simu}

Due to the limited angular resolution of Pan-STARRS, we focused on the search for wide-separation lens systems with bright
arcs that can be easily recognized by eye. We imposed a lower limit on the Einstein radius $\theta_{\rm E}$ of mocks of
1.5\arcsec, larger than the median FWHM of PS1 seeing in $gri$ bands. This ensured that individual counter-images are well
deblended from each other in the mocks (albeit often blended with the lens). Each lens deflector drawn from the LRG catalog was
cross-matched with a random COSMOS source at $z_{\rm source} > z_{\rm deflector}$, rejecting pairs with $\theta_{\rm E} < 1.5$\arcsec
\footnote{Calculated given the known lens redshift and velocity dispersion and the known source redshift.}, and repeating
the process iteratively to obtain 90\,000 lens+source pairs. Focusing on larger Einstein radii amounts to selecting LRGs in
the high-mass range, with $v_{\rm disp} \sim 230$--400~km~s$^{-1}$, and redshifts $z_{\rm d} \sim 0.1$--0.7 representative of
the input BOSS sample, and sources in the redshift range $z_{\rm s} \sim 0.5$--3.0. The pairs mainly cover the $\theta_{\rm E}$
range of 1.5--3.0\arcsec, which is dominated by galaxy-scale dark-matter lens halos on the high-end of the mass distribution
\citep{oguri06}, while group-scale lenses contribute predominantly for image separations above 3\arcsec. $\theta_{\rm E}$
values were not uniformly distributed but dropped by a factor 100 from 1.5\arcsec to 3.0\arcsec, akin to real galaxy-scale
lenses, implying that our CNN is predominantly exposed to mock systems with $\theta_{\rm E} \sim 1.5$\arcsec.

For each pair, mock images were created with the simulation pipeline described in Schuldt et al. (in prep.). In short, the
lens potential was modeled with a Singular Isothermal Ellipsoid (SIE) profile, based on the known $v_{\rm disp}$ and $z_{\rm d}$,
and using the centroid, axis ratio, and position angle from the $i$-band light distribution, with random perturbations
typical of SLACS lenses \citep{bolton08}. The combined HSC+F814W cutouts of COSMOS sources were randomly positioned in the
source plane, over regions next to the caustics corresponding to magnifications $\mu \geq 5$. The sources were then lensed onto
the image plane with the {\tt GLEE} software \citep{suyu10,suyu12}. The resulting frames were convolved with the Pan-STARRS PSF
model described below, resampled and rescaled using Pan-STARRS zero-points, and eventually coadded with the lens LRG cutouts to
obtain the final mock image. The process was repeated for $gri$ bands. In order to produce a set of mocks with systematically
bright lensing features, we artificially boosted the lensed source brightness by one magnitude in all bands. Mocks with faint
arcs were placed iteratively closer to the caustics to ensure all lensed sources have ${\rm S/N > 10}$ in $i$
band\footnote{Between the peak lensed source flux and the local background rms level.}.

The Pan-STARRS analysis pipeline computes PSF models at individual positions of stack images over a grid of about 8\arcmin\
steps, and interpolates these models to predict the PSF FWHM across the sky \citep{magnier16}. This introduces deviations
between the modeled and true PSF, since the latter varies on very small scales due to stacking of individual exposures with
variable FWHM. However, we found that these deviations are usually within 10\% when comparing the tabulated FWHMs with those
measured on isolated, unsaturated stars from GSC-DR2 \citep{lasker08}. We therefore created a library of $gri$ PSF models
in steps of 0.05\arcsec\ FWHM, by stacking PS1 postage stamps of nine to 11 stars with adequate PSF FWHM. For each mock,
lensed arcs were convolved with the PSF model corresponding to the PSF FWHM listed on PS1 tables at the position of the lens
LRG.

We generated a total of 90\,000 mock lens systems (see Fig.~\ref{fig:simu}). Lens LRGs selected multiple times were rotated by
$k \pi/2$ and used only once for a given orientation, with different lensed arc configurations, so the networks never got the
exact same image several times as input. The mocks cover realistic source colors and lensing configurations, including quads, 
near-complete Einstein rings, fold and cusp arcs, and doubles (see Schuldt et al., in prep.). Constraining the source plane 
positions to large magnifications likely biases our set to lower fractions of doubles than in blind samples of real lenses.

\subsection{Photometry of mock lens systems}
\label{ssec:photo}
      
Focusing the CNN lens search on a subset of the three billion sources detected in the PS1 3$\pi$ survey requires a preselection
of sources based on their properties released in public catalogs. For this purpose, we computed the photometry
of mock lensed systems in the same way as the PS1 image processing pipeline \citep{magnier16}. Fixed aperture photometry
is particularly important to measure reliable colors. We derived the integrated magnitudes of our mocks within the four
smaller PS1 circular apertures of 1.04\arcsec (R3), 1.76\arcsec (R4), 3.00\arcsec (R5), and 4.64\arcsec\ (R6) radii, which
are best suited to capture color gradients due to the presence of lensed arcs at angular separations of 1.5--3.0\arcsec\
with respect to the lens center. The two largest of these apertures are also expected to be relatively good proxies of
the integrated magnitudes of mocks. We used SExtractor in dual-image mode, with a 3$\sigma$ detection threshold in r-band, 
and assuming an ideal sky subtraction in PS1 deep stacks \citep[see details in][]{waters16}. To compute the aperture 
magnitudes of a given mock, the image zero-points were taken from the PS1 catalog of stack detections at the position of
the LRG used to produce this mock.

The method was tested on LRG-only images by comparing SExtractor estimates with those from the PS1 catalog, for standard
stacks. For the four apertures, fitting the distributions of magnitude offsets between these two estimates with Gaussian
functions led to $\mu = 0.00$--0.02 and $\sigma = 0.02$--0.05 in $g$ band, $\mu = 0.00$--0.01 and $\sigma = 0.01$--0.02
in $r$ band, $\mu = 0.00$ and $\sigma = 0.01$--0.02 in $i$ band, which proves the overall robustness of our photometry.
Unsurprisingly, the scatter only rises above these average values for the large aperture magnitudes of fainter objects,
which mostly enclose background noise. These residual biases up to 0.1--0.2~mag for $\gtrsim$22.0 mag likely indicate
small differences in the local background subtraction between both methods, or the contamination from neighbors which
were subtracted by the PS1 pipeline \citep{magnier16} but not with SExtractor. Nonetheless, these offsets remain minor
compared to other uncertainties in the analysis. On the contrary, Kron, Petrosian and Sersic photometry were discarded
due to systematic biases with respect to values tabulated in PS1 catalogs.

\section{Systematic search of strong lenses}
\label{sec:cnn}

The next sections describe the steps followed for the generic lens search on the full Pan-STARRS 3$\pi$ survey.

\subsection{Preselecting Pan-STARRS detections}
\label{ssec:presel}

As shown in Fig.~\ref{fig:catsearch}, the 90\,000 mock lens systems have globally bluer colors than the LRG sample due to
the relative color of lensed arcs, and they are brighter than $\sim$70\% of sources detected in the PS1 stack images.
This population of fainter and bluer PS1 galaxies can be excluded from the analysis. We used simple color-magnitude cuts in
the $(g-i)$ vs. $i$, $(g-r)$ vs. $r$, and $(r-i)$ vs. $i$ diagrams for the R3, R4, R5, R6 circular apertures to rule out
regions in these diagrams that are not representative of the mocks (i.e., not colored in blue in Fig.~\ref{fig:catsearch}).
These cuts are conservative and include 96\% of the mocks, according to the aperture magnitudes obtained in
Sect.~\ref{ssec:photo}, while excluding $\sim$84\% of PS1 sources. They were applied to the complete catalog of stack
detections from PS1 DR2 using the PanSTARRS1 Catalog Archive Server Jobs System\footnote{http://casjobs.sdss.org/CasJobs},
with ${\rm {\tt detectionFlags3} > 65536}$ to remove multiple entries and to select detections from the optimal stack image
\citep{flewelling16}.

The stars were removed from the resulting sample using the $r$-band cuts from \citet{farrow14}:

\noindent $r_{\rm Kron} - r_{\rm PSF} < -0.192+0.120 \times (r_{\rm Kron}-21) + 0.018 \times (r_{\rm Kron}-21)^2$.

This selection conservatively included 98\% galaxies, those discarded being mainly at the faint end ($r \gtrsim 21$) and
with higher magnitudes than our mocks. While these cuts misidentified saturated stars with $r \lesssim 14$ as galaxies
\citep{farrow14}, such bright sources had already been excluded from the analysis.

Regions with elevated Galactic dust extinction were removed as strong reddening could alter our selection on the catalog
level. The interstellar dust reddening 2D map of \citet{schlegel98} were loaded with the {\tt dustmaps} python interface
from \citet{green18}. After converting to PS1 bandpasses using coefficients in Table 6 of \citet{schlafly11}, we applied a
reddening threshold of ${\rm E(}g-i{\rm)} < 0.3$. These steps resulted in a catalog of 23.1 million galaxies for classification.

\subsection{Applying machine learning to the Pan-STARRS photometry}
\label{ssec:catsearch}

Limitations due to download speed of PS1 cutouts from the archive can be overcome by further reducing the size of this
catalog with additional selection criteria. The simple color and magnitude cuts applied to the complete PS1 catalog do not
capture all photometric properties of mock lens systems, such as their precise locus on two-dimensional color-color and
color-magnitude diagrams or their radial color gradients. For instance, Figure~\ref{fig:catsearch} indicates that mocks
are generally redder within the smaller R3 aperture of 1.04\arcsec\ radius than within external, concentric annuli between
R3 and R4, and between R4 and R5. These gradients are caused by the presence of bluer, lensed arcs at $>$1.5\arcsec\ from
the lens center and disappear on the LRG-only sample. To exploit this information, we trained a simple fully-connected
neural network on the photometry of mocks and random PS1 sources using aperture fluxes that ensure robust colors.

The data set contained $gri$ fluxes in the four apertures for 90\,000 lens and 90\,000 nonlens examples as inputs, and the
ground truth labels of 1.0 and 0.0, respectively, as outputs for binary classification. The negative examples were fluxes of
random sources that matched our loose color and magnitude $gri$ cuts in Sect.~\ref{ssec:presel}. The data set was split into
training, validation, and test sets with respective fractions of 56\%, 14\%, and 30\%. All fluxes were normalized to the average
over the entire data set in order to speed up the learning process. The network architecture consisted of 12 dimensional input
data, three fully connected hidden layers of 50, 30, and five neurons each, with Rectified Linear Unit \citep[ReLU,][]{nair10}
nonlinear activations\footnote{${\rm ReLU(x) = max(x,0)}$}, and a single-neuron output layer with sigmoid
activation\footnote{${\rm sigmoid(}x{\rm)=\frac{1}{(1+e^{-x})}}$}.

During the training phase, the network derives a model for classifying galaxies in the training set as lenses or nonlenses
according to their input photometry, as briefly summarized with the following stages. After the weight parameters and bias
in each neuron are initialized, a subset of the training data is passed through the entire network to calculate predicted
labels (forward propagation), and the difference between predictions and ground truth labels is quantified with a loss
function $L$. This information is propagated to the network weights and biases \citep[back propagation,][]{rumelhart86} which
are then modified using a gradient descent algorithm to minimize the total loss and improve the model. These stages are
repeated iteratively to perform a complete pass through the entire training set, corresponding to one epoch, and then over
multiple epochs until the model reaches optimal accuracy. After each epoch, the validation loss is evaluated by classifying
inputs from the validation set, in order to determine whether the decrease in training loss reveals better performance or an
overfitting to the training set. After training, the network performance is finally quantified using independent data from
the test set. Further details can be found in the review of \citet{lecun15}.

Network parameter optimization is performed via mini-batch stochastic gradient descent, a common variant that consists of
splitting the training set into small batches and adjusting the weights according to the average corrections over each batch.
Our network minimized the cross-entropy loss function which penalizes robust and incorrect predictions, and is expressed as
follows for a binary classification problem 
\begin{equation}
L(y,p) = - \frac{1}{N} \sum_{i=0}^{N} y_{i} \log(p_{i})+(1-y_{i}) \log(1-p_{i})
\end{equation}
where $y_i$ are the ground truth labels, and $p_i$ the network predictions, namely scores in the range $[0,1]$ resulting from
the sigmoid activation on the output layer. The loss was computed over each batch of size $N$. To avoid unbalanced splits of
the data set, we used 5-fold cross-validation that consists of reshuffling the training and validation sets and building the
performance metrics. Cross-validation runs trained over 500 epochs were used to optimize the neural network hyperparameters
with a grid search, varying the learning rate over the range $[0.0001,0.1]$ and the weight decay over $[0.00001,0.01]$, with
momentum fixed to 0.9. We trained a final network with the entire data set using an optimal learning rate and weight decay of
0.001 and 0.00001, respectively, and applying early stopping at epoch 193 that matched the lowest average loss over the
cross-validation runs. 

Among the 23.1 million input galaxies, 1\,050\,207 were assigned scores $p_{\rm cat} > 0.5$ indicating that their $gri$ aperture
photometry is consistent with the mocks. Fig.~\ref{fig:catsearch} illustrates the network predictions by comparing the colors
and magnitudes of random galaxies, mocks, and galaxies having $p_{\rm cat} > 0.5$. The good agreement between 2$\sigma$ contours
of mocks and $p_{\rm cat} > 0.5$ shows that the network correctly captures the position of mocks in color-magnitude diagrams, as
well as their color variations within different apertures. Moreover, our photometric selection was successfully tested on the
aperture fluxes of known strong lensing systems listed as grade A or B with $\theta_{\rm E} = 1.0$--3.0\arcsec\ in the MasterLens
database\footnote{http://admin.masterlens.org}, and with lensed arcs visible in the PS1 stack images. We therefore kept these
1\,050\,207 galaxies for CNN classification.

\begin{figure}
\centering
\includegraphics[width=.5\textwidth]{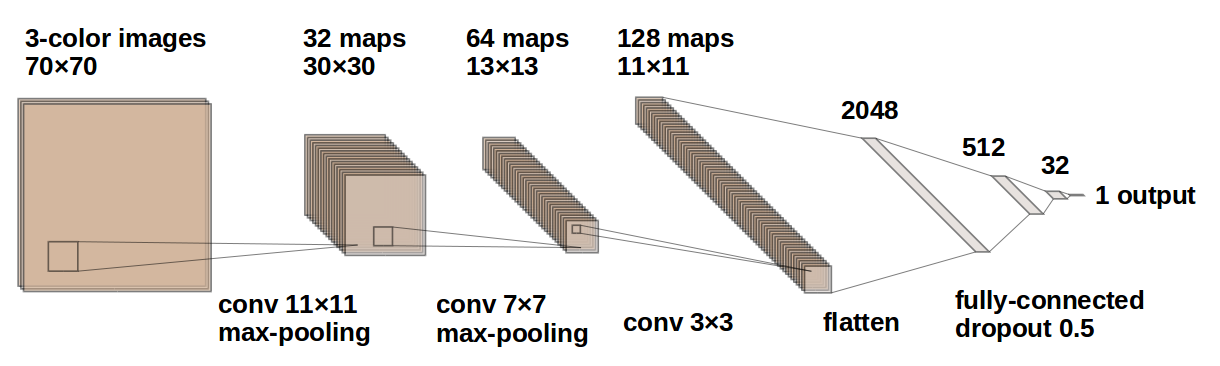}
\caption{
  Architecture of the convolutional neural network, inspired from LeNet \citep{lecun98}, and comprised of three convolutional layers
  with 11$\times$11, 7$\times$7, and 3$\times$3 kernel sizes, and 32, 64, and 128 filters, respectively, followed by three fully
  connected hidden layers with 2048, 512, and 32 neurons. ReLU activations were applied between each layers. Max-pooling layers with
  2$\times$2 kernel sizes and stride = 2 were inserted after the first two convolutional layers, and dropout of 0.5 was used before
  the fully connected layers.}
\label{fig:archi}
\end{figure}

\subsection{Training the convolutional neural networks}
\label{ssec:training}

\begin{figure}
\centering
\includegraphics[width=.5\textwidth]{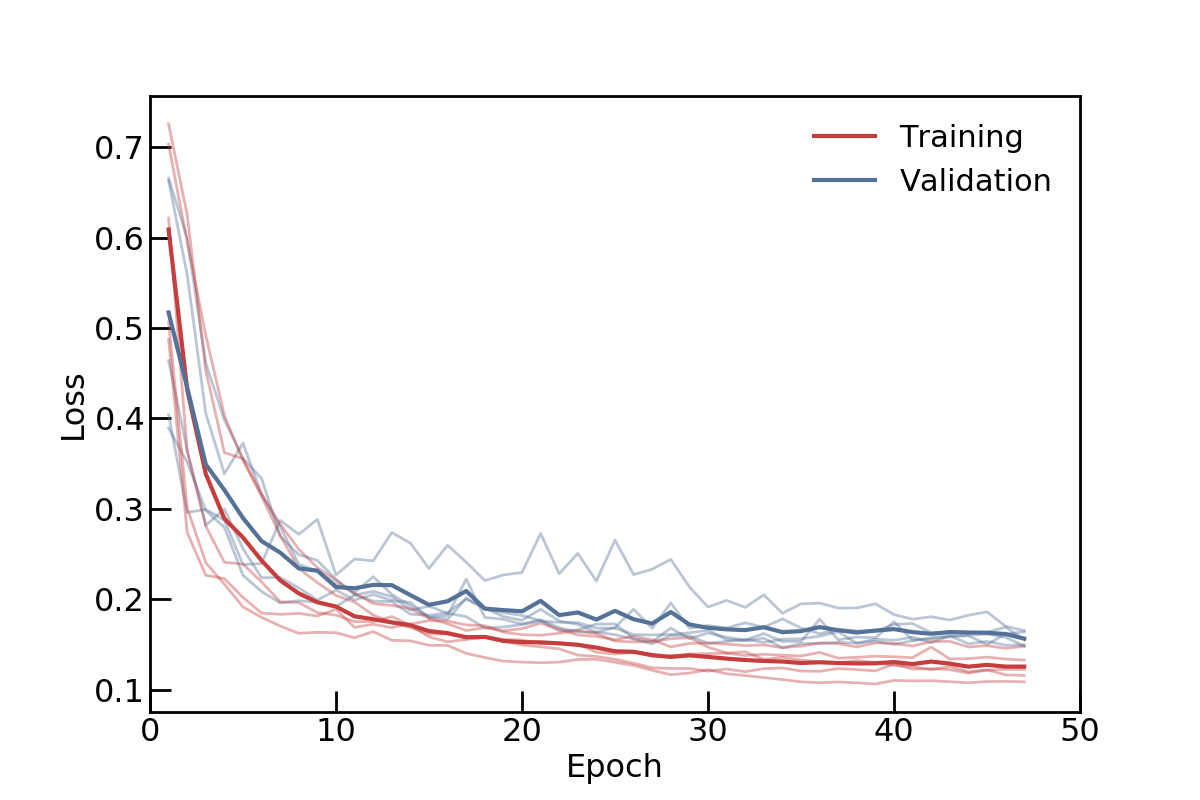}
\caption{
  Training of the CNN with optimized hyperparameters and using early stopping. The training loss (red curve) and validation loss
  (blue curve) were taken as the average of all cross-validation runs (light red and blue curves).}
\label{fig:training}
\end{figure}

\begin{figure}
\centering
\includegraphics[width=.5\textwidth]{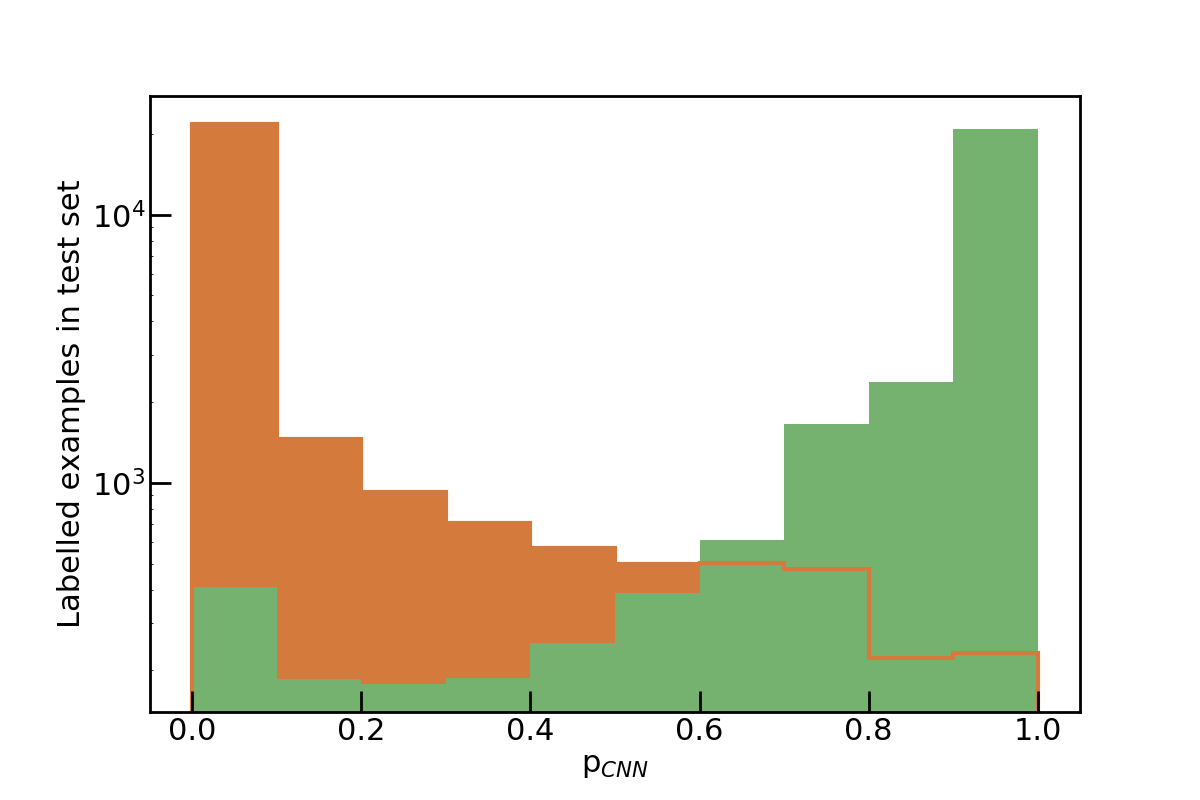}
\caption{
  Distribution of network predictions compared with the ground-truth for lenses (green) and nonlenses (orange) in the test set.}
\label{fig:testing}
\end{figure}

Data sets for the CNNs included 180\,000 images in $g$, $r$, and $i$ bands with the same fraction of positive
(lens) and negative (nonlens) examples. Positive examples were taken from the sample of simulated lens galaxies with
$\theta_{\rm E} > 1.5$\arcsec\ described in Sect.~\ref{ssec:simu}. The choice of negative examples is strongly influencing the
network predictions and was modified iteratively to improve the network performances (see Sect.~\ref{ssec:perf}). In short, we
boosted the fraction of galaxies with specific morphological types using incorrect identifications of strong lenses in previous
networks. This allowed the network to learn how to discriminate strongly lensed arcs from the usual contaminants such as extended
arms of low redshift spirals, lenticular galaxies, and mergers, and to distinguish isolated LRGs from LRGs with the relevant
strong lensing features depicted in the mocks. Our resulting set of 90\,000 negative examples included:
\begin{itemize}
\item 30\% LRGs selected directly from the catalog of SDSS LRGs on the high-end of the mass distribution used to create
  the mocks,
\item 20\% spirals classified as likely face-on galaxies in Galaxy Zoo 2 \citep{willett13} and with $r$-band Sersic radii
  $<$4.5\arcsec\ from PS1 \citep{flewelling16}, to restrict to blue spiral arms with similar extension as the lensing
  features present in the mocks,
\item 10\% smooth, isolated galaxies from Galaxy Zoo 2 without bright companion and bluer colors than LRGs,
\item $<$1\% galaxies with apparent dust lanes identified in Galaxy Zoo 2 \citep{willett13},
\item 32\% randomly selected galaxies from the PS1 catalog, including diverse types, groups and mergers, and with negligible
  contamination from the rare strong lenses,
\item 7\% false positives from previous neural networks selected by visually classifying candidates with scores $>$ 0.9.
\end{itemize}

Negative examples were not included in the $p_{\rm cat}>0.5$ sample to be classified with the CNN, but they broadly followed the
same color and magnitude distributions in $gri$ bands. Some examples are shown in Fig.~\ref{fig:simu}. The data set was split
into training, validation, and test sets with the same proportions as before (Sect.~\ref{ssec:catsearch}).

We used data augmentation and applied random shifts between --5 and +5 pixels to all images so the network becomes invariant on
small positional offsets. This resulted in input images of $70\times70$ pixels which conservatively included all emission from
the central galaxy of interest. All survey galaxies in the set of negative examples were unique, and LRGs used multiple times
in the sample of mocks were rotated by $k\pi/2$ so they appeared only once with a given orientation in the set of positive
examples. This ensured that the network was not sensitive to preferential orientations. Other common image augmentation
techniques, such as image stretching, normalization and rescaling \citep[e.g.,][]{petrillo17,petrillo19a} were discarded as they
did not significantly improve the learning process.

The CNN architecture was inspired from LeNet \citep{lecun98} and from the lens modeling CNN of Schuldt et al. (in prep.). After
the $70\times70\times3$ input layer, it contains three convolutional layers with $11\times11$, $7\times7$, and $3\times3$
kernel sizes, and 32, 64, and 128 filters, respectively, followed by three fully connected hidden layers with 2048, 512, and
32 neurons (see Fig.~\ref{fig:archi}). ReLU activations act as nonlinear transformations between each one of these layers.
Max-pooling layers \citep{ranzato07} with $2\times2$ kernel sizes and ${\rm stride=2}$ are inserted after the first two
convolutional layers and are essential to make the CNN invariant to local translations of the relevant features in $gri$
image cutouts, while reducing the network parameters. Dropout regularization \citep{srivastava14} with a dropout rate of 0.5
is applied before the fully connected layers. This is an efficient regularization method that consists of randomly ignoring
neurons during training in order to reduce overfitting on the training set and improve the CNN generalization. The output
layer consists of a single neuron with sigmoid activation and results in a score, $p_{\rm CNN}$, in range $[0,1]$ which corresponds
to the network lens or nonlens prediction\footnote{For uncalibrated networks, this score differs from the likelihood of correct
  classification.}. Our CNN with moderate depth is well suited for binary classification of small PS1 cutouts. 

During the training process, the CNN learns the relevant patterns in $gri$ images by adjusting the convolutional kernel
weights, through a minimization of the binary cross-entropy loss between ground truth and predicted labels. After the
gradient calculation and optimization, information learned by the network is stored in the two-dimensional filters. As for
the catalog-level network, we used mini-batch gradient descent with a batch size of 128 and performed five cross-validation
runs. We found an optimal learning rate and weight decay of 0.0006 and 0.001, respectively, using a grid search with momentum
fixed to 0.9. The number of training epochs was then chosen from the minimum average validation loss over the cross-validation
runs, which corresponded to optimal network performance without overfitting.

\begin{figure}
\centering
\includegraphics[width=.5\textwidth]{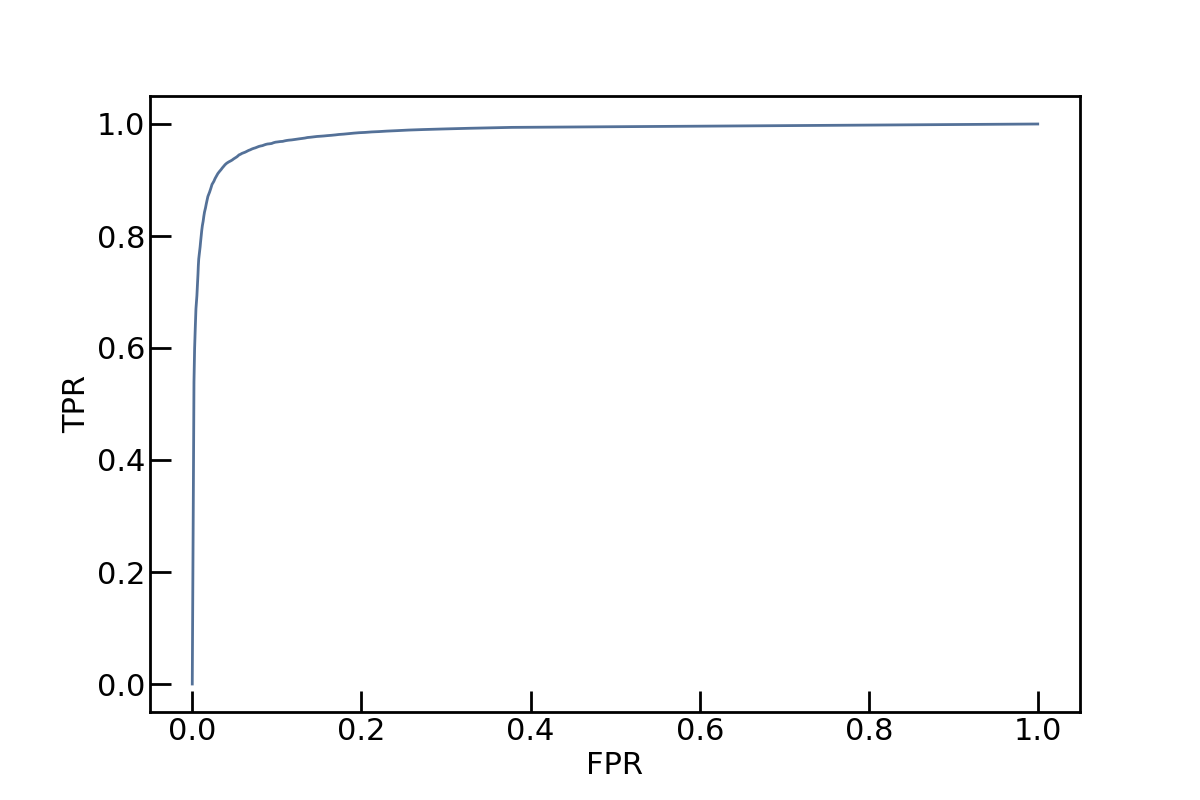}
\caption{
  Receiver Operating Characteristic curve for the trained CNN showing the true position rate (TPR) as function of the false
  positive rate (FPR) for different lens identification thresholds. The corresponding area under curve (AUC) is 0.985.}
\label{fig:roc}
\end{figure}

The evolution of the training and validation loss for the network with optimized hyperparameters is shown in
Fig.~\ref{fig:training}, until epoch 47 which corresponds to the lowest validation loss. The gap between both curves (generalization
gap) is small, showing that the model predictions do not deteriorate much on new data with similar properties as the training set.
The final network performances were characterized with the test set which was not seen during training and validation and contains about
54\,000 entries. In Fig.~\ref{fig:testing}, we show the model probability predictions for all lens and nonlens examples in the test
set. Lenses dominate the distribution for $p_{\rm CNN} > 0.6$. The model reaches 94.2\% accuracy, 93.1\% purity and 95.5\% completeness
on this set suggesting good pattern recognition abilities on new images\footnote{Accuracy was defined as the sum of true positives and
  true negatives over the total number of systems (lenses + nonlenses), and purity was defined as the number of lenses correctly
  identified with $p_{\rm CNN}>0.5$ over the number of systems with $p_{\rm CNN}>0.5$.}. In addition, the Receiver Operating Characteristic
(ROC) curve in Fig.~\ref{fig:roc} illustrates the relation between the true positive rate (TPR, the number of lenses correctly
identified over the total number of lenses) and the false positive rate (FPR, the number of nonlenses identified as lenses over the
total number of nonlenses) in our trained model. The curve was obtained by varying the network probability threshold for lens
identification in the range $[0,1]$, and an ideal network would correspond to an area under curve (AUC) of 1. Our CNN gives
${\rm AUC} = 0.985$, and can correctly identify 95.5\% of lenses in the test set with a FPR of 7.1\% using $p_{\rm CNN} > 0.5$, or
77.0\%  of lenses in the test set with a FPR of 0.8\% using $p_{\rm CNN} > 0.9$. After optimizing hyperparameters and quantifying the
performance of the network on the test set, we trained a final CNN with the complete data set of 180\,000 labeled images and we fixed
the resulting model parameters to classify the 1\,050\,207 galaxies with $p_{\rm cat} > 0.5$.

\begin{figure*}
\centering
\includegraphics[width=.98\textwidth]{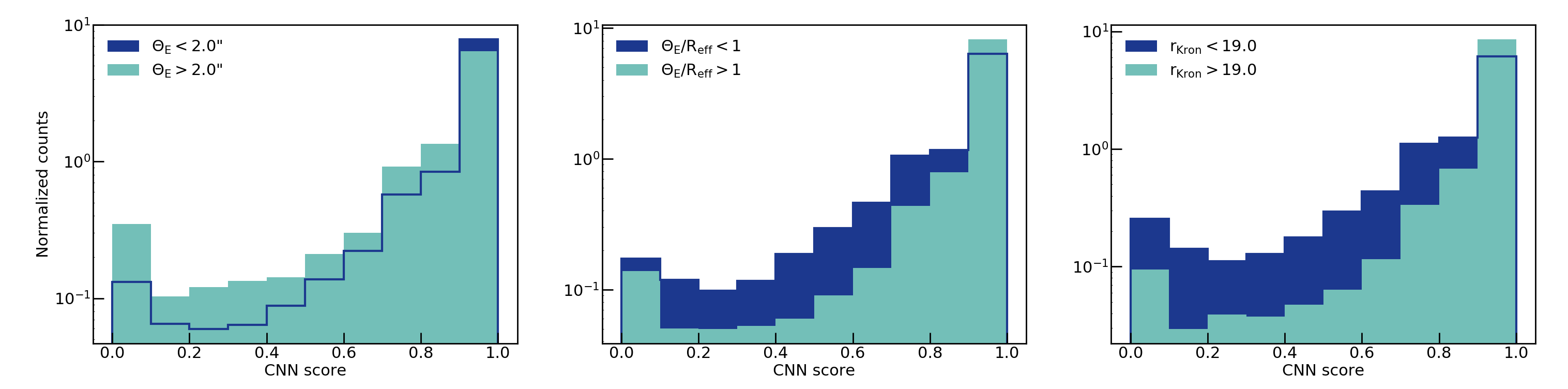}
\caption{
  Normalized distributions of CNN scores for mocks lenses included as positive examples in the test set, for different ranges of
  Einstein radii (left), $\theta_{\rm E}/R_{\rm eff}$ (middle), and $r_{\rm Kron}$ (right). }
\label{fig:testprop}
\end{figure*}

\subsection{Evaluating the network performance}
\label{ssec:testing}

\begin{figure}
\centering
\includegraphics[height=.45\textwidth]{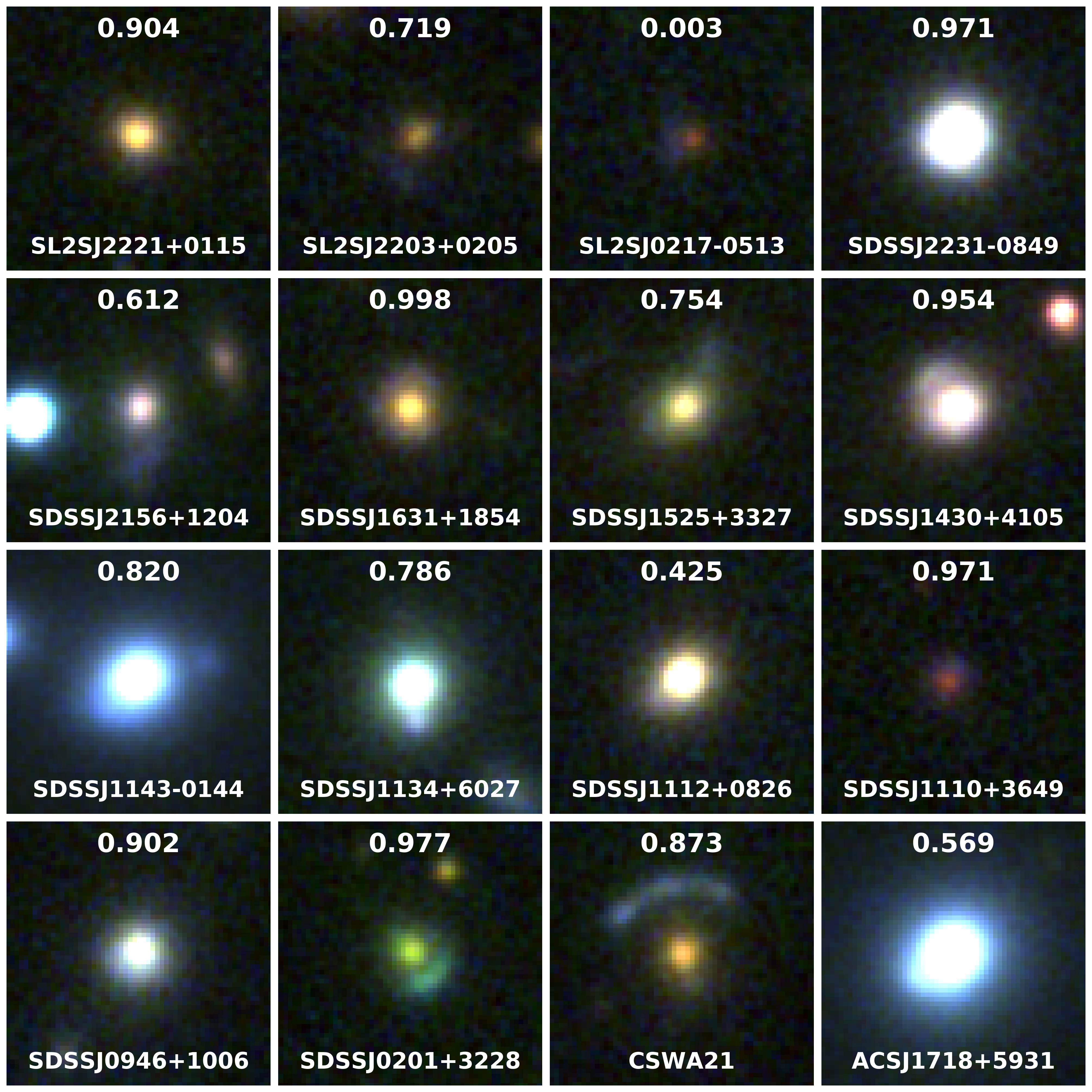}
\caption{
  Three-color images of the 16 confirmed lens systems in the MasterLens database that have clear strong lensing signatures in
  Pan-STARRS images, and Einstein radii of 1--3\arcsec\ similar to our mocks. The CNN scores are displayed at the top of each panel.
  Images are 15\arcsec~$\times$~15\arcsec.}
\label{fig:testnet}
\end{figure}

The performance of the network was further evaluated on galaxies with various properties on the test set. Fig.~\ref{fig:testprop}
depicts the normalized distributions of scores for positive examples in different bins of $\theta_{\rm E}$, $\theta_{\rm E}/R_{\rm eff}$,
and $r_{\rm Kron}$, where $R_{\rm eff}$ is the effective radius from the $r$-band Sersic fit of the LRG-only image \citep{flewelling16}.
Somewhat counterintuitively, scores are closer to 1.0 for lenses with $\theta_{\rm E} < 2$\arcsec. The low fraction of
$\theta_{\rm E} > 2$\arcsec\ in the training set might explain the slightly lower CNN performance on these systems, which generally
do not have the most challenging morphologies. Histograms for $\theta_{\rm E}/R_{\rm eff}$ demonstrate the ability of the network to
assign $p_{\rm CNN}$ closer to 1.0 for mocks with Einstein radius larger than the effective radius of the lens light distribution,
where lensed arcs are in principle better deblended from the lens. We also find a higher fraction of scores $p_{\rm CNN}>0.9$ for
lens LRGs with higher $r_{\rm Kron}$ (i.e., fainter LRGs), perhaps because the brightest lenses outshine the lensed source emission.
Nonetheless, these variations remain generally minor.

Finding acceptable network performances on the test set might be misleading as it relies on our choices in simulating the
strong lenses and assembling a set of negative examples. A valuable independent test consists of applying the CNN to strong
lenses from the literature. For that purpose, we collected all grade A and B galaxy-scale lenses in the MasterLens catalog,
restricting to Einstein radii $\sim$1--3\arcsec\ similar to the range probed by our 90\,000 mocks. These systems were discovered
from various techniques including the identification of emission lines from star-forming galaxies behind LRGs using
spatially-integrated spectra \citep[SLACS,][]{bolton08}, and the analysis of high-quality imaging from HST
\citep[e.g., in COSMOS,][]{faure08} or deep multiband surveys \citep[e.g., CFHTLS SL2S,][]{cabanac07,gavazzi14,more16}. Most
of these lenses are not detectable in $gri$ Pan-STARRS stacks and need to be excluded from our test set. We thoroughly scanned
all PS1 $gri$ single-band and 3-color images of this sample to find those with detected lensed arcs, and assembled a test set
of 16 systems. While these published lenses have colors, $z_{\rm d}$, $z_{\rm s}$, and configurations similar to our mocks, some
of their multiple images are strongly blended with the lenses and difficult to identify with PS1 data.

Pan-STARRS cutouts of these 16 lenses were scored with our trained neural network and the results are presented in
Fig.~\ref{fig:testnet}. A total of 14/16 lenses are correctly identified as $p_{\rm CNN} > 0.5$ by the CNN, while 9/16 and 7/16 have
higher scores $p_{\rm CNN} > 0.8$ and $p_{\rm CNN} > 0.9$, respectively. SL2SJ0217$-$0513 and SDSSJ1112$+$0826 are the two incorrect
classifications. SL2SJ0217$-$0513 has a faint lens galaxy falling in the upper range of the redshift distribution ($z_{\rm d}=0.646$)
and blended with a faint blue arclet. It is worth noting that the network performs better for SDSSJ1110$+$3649 ($p_{\rm CNN} = 0.971$)
which has a similar morphology to SL2SJ0217$-$0513, but with the addition of a low S/N, blue counter image on the other side of the
lens galaxy. SDSSJ1112$+$0826 has typical properties of our mocks but lacks a bright counter image. On the other hand, lenses with
well-detected and deblended arcs and counter images (e.g., SDSSJ1430$+$4105, SDSSJ0201$+$3228, CSWA21) have high scores, showing
that the network is able to extract relevant features. 

Our comparison extends to systems with Einstein radii below the 1.5\arcsec\ cutoff applied to our simulations. The CNN performance
in this regime remains acceptable, as illustrated by the scores of 0.786, 0.612, and 0.971 assigned respectively to SDSSJ1134$+$6027,
SDSSJ2156$+$1204, SDSSJ2231$-$0849 that have $\theta_{\rm E} \sim 1.1$\arcsec. In contrast, with $\theta_{\rm E} = 3.26$\arcsec\ due
to its group-scale environment \citep{auger13}, CSWA21 falls on the upper range of the Einstein radius distribution that is
underrepresented with only a few hundred examples in our training set. This system is nonetheless given a score of 0.873 and
confirms that the CNN can identify these simple configurations. Interestingly, the four systems with $\geq 4$ magnified images
according to the MasterLens database have $p_{\rm CNN} > 0.9$. The small number of test lenses however prevents robust estimates
of the method purity and completeness.

\subsection{Discussion on the CNN training}
\label{ssec:perf}

The final version of the CNN from Sect.~\ref{ssec:training} was selected from a range of networks with different architectures,
after testing the impact of the training set content. To identify the optimal network, we compared scores assigned to the 16
known test lenses and required low false positive rates by examining $gri$ cutouts of the few hundred galaxies with highest
scores $p_{\rm CNN}$. Overall, different choices of positive and negative examples had much stronger impact, inducing variations
in the number of good lens candidates from visual inspection by a factor of $\gtrsim$10 due to the network learning different
features, while changes in the CNN architecture only offered slight improvements.

The sets of negative examples tested include: (1) random PS1 sources drawn from the preselection in Sect.~\ref{ssec:presel};
(2) typical LRGs selected as in \citet{eisenstein01}, mostly less massive than LRGs in mocks; (3) high-mass LRGs similar
to those used in Sect.~\ref{sec:simu}; (4) a combination of LRGs, face-on spirals, and random sources (varying the fractions
of LRGs and contaminants). Scores on the MasterLens systems from CNNs using sets (1) and (4) were comparable to those in
Fig.~\ref{fig:testnet}, but introduced an overwhelming number of $\gtrsim$400\,000 galaxies with $p_{\rm CNN}>0.5$ and $\gtrsim$250\,000
with $p_{\rm CNN}>0.9$ implying high false positive rates, and were ruled out. Other sets of positive examples were tested by (1)
modifying the $\theta_{\rm E}$ lower limit, and (2) suppressing the artificial boost in arc brightness to get more realistic
lens over source flux ratios. Both CNNs trained on fainter arcs, more strongly blended with the lens significantly reduced the
fraction of genuine lens examples with scores $p_{\rm CNN} \sim 1$ and were also discarded.

Our tests on the architecture include: (1) adding and removing one convolutional layer, or one fully-connected layer; (2)
changing the number of neurons, the kernel sizes or number of feature maps in these layers; (3) using smaller or larger strides
on the max-pooling layers; (4) modifying the dropout rates; (5) implementing batch normalization \citep{ioffe15}. Each of these
changes degraded the network performance as measured from the loss and ROC curves on the test set, and from the 16 MasterLens
systems. Using dropout normalization before fully connected layers (as in Fig.~\ref{fig:archi}) turned out to be the most
efficient solution to reduce overfitting.

\subsection{Visual classification}
\label{ssec:visu}

\begin{figure*}
\centering
\includegraphics[width=.98\textwidth]{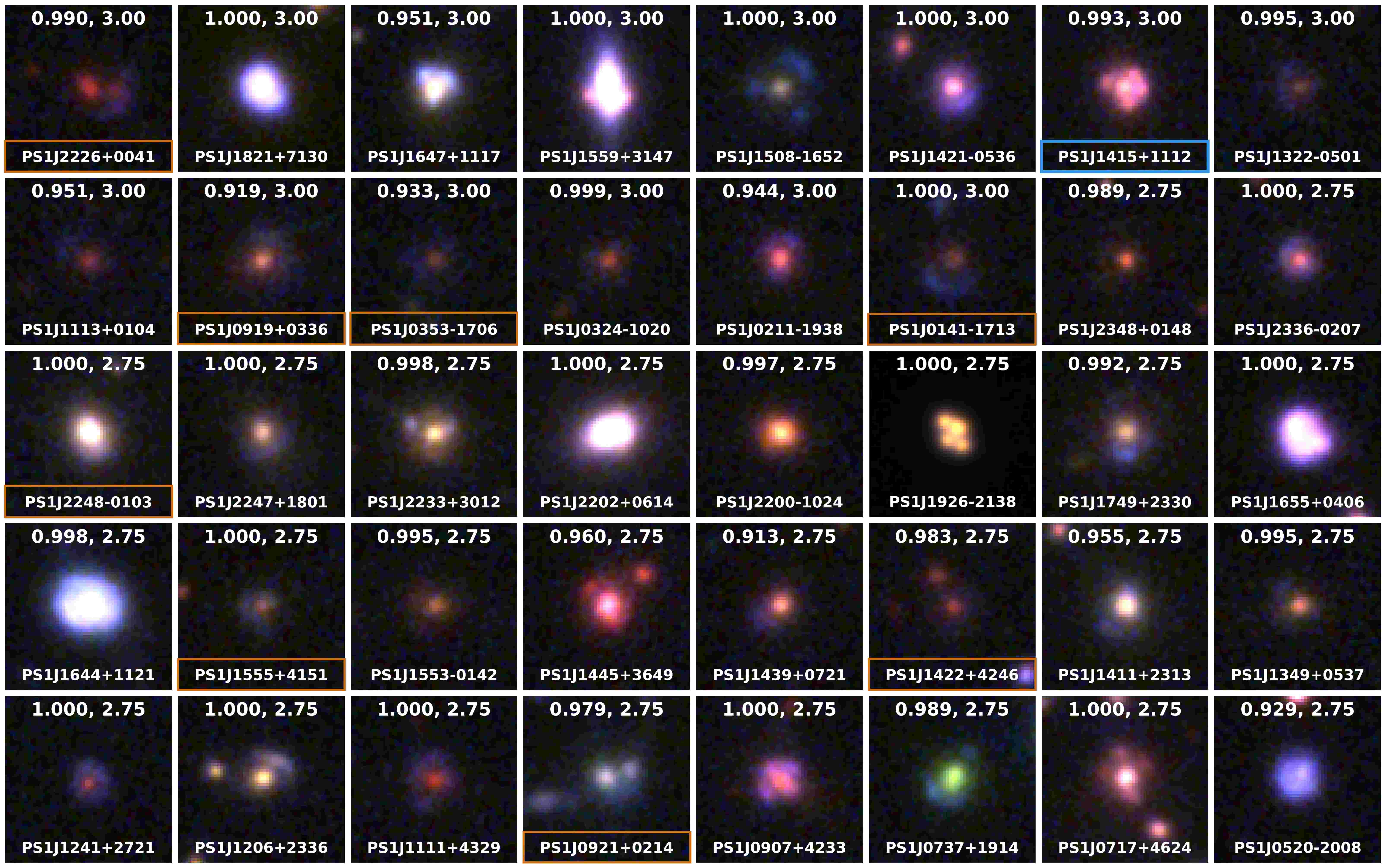}

\vspace{3mm}\includegraphics[width=.98\textwidth]{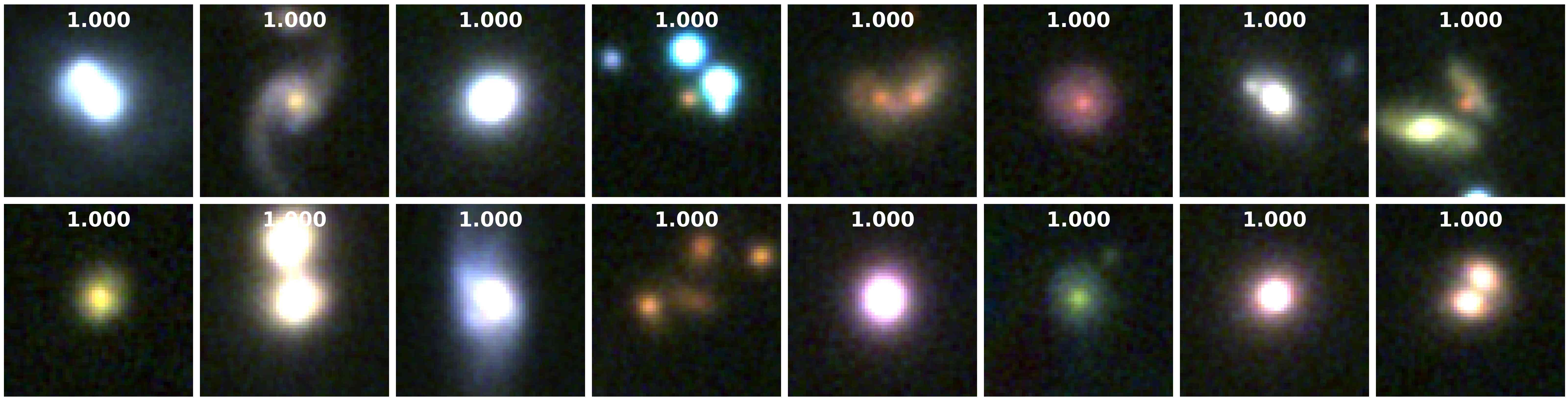}
\caption{
  {\it Top:} Pan-STARRS 3-color $gri$ images of a subset of candidates with grades $\geq 2$ from visual inspection of CNN scores
  $p_{\rm CNN} > 0.9$ (the complete figure is available in Appendix A). Numbers displayed on top of each panel are CNN scores
  ($p_{\rm CNN}$, left) and average visual grades (right). Candidates with PS1 names marked in orange have been previously published
  in the literature (see Table~\ref{tab:cand} and Sect.~\ref{ssec:cnncand}) Those marked in blue show unambiguous spectral signatures
  of high-redshift background sources in our inspection of SDSS BOSS DR16 data. {\it Bottom:} Examples of random false positives with
  $p_{\rm CNN} = 1$. All cutouts are 15\arcsec~$\times$~15\arcsec.}
\label{fig:cnncand}
\end{figure*}

Galaxies with high CNN scores, $p_{\rm CNN}$, were visually inspected by different authors to assign a final grade. We started
classifying galaxies with $p_{\rm CNN}$ close to 1.0, and progressively lowered the threshold to introduce additional galaxies
until the fraction of reliable candidates from visual inspection became too low.

We graded the single-band and 3-color PS1 cutouts in $gri$ bands, zoomed to 12\arcsec~$\times$~12\arcsec, and optimally displayed
with linear and arcsinh scaling using dedicated scripts to emphasize faint, sometimes blended strong lensing features\footnote{code
  adapted from https://github.com/esavary/Visualisation-tool}. To aid the visual classification, we plotted $grz$ single-band and
3-color postage stamps from the DESI Legacy Imaging Surveys that significantly overlaps the PS1 3$\pi$ survey footprint on the
extragalactic Northern sky, and that provides slightly deeper, higher quality images \citep{dey19}, and we plotted residual frames
from subtraction of the best-fit light profile. On smaller regions of the sky, we also included $gri$ images from HSC DR2
wide-field surveys \citep{aihara19}. The set of CNN candidates was divided into four equal parts, each inspected either by R.\,C.,
S.\,S., S.\,H.\,S., or S.\,T., in order to assign one of the following grades: {\tt 0}: nonlens, {\tt 1}: maybe a lens,
{\tt 2}: probable lens, {\tt 3}: definite lens, similarly to \citet{sonnenfeld18} and \citet{jacobs19b}. All candidates with
grades $\geq 2$ from this first iteration were then inspected by the three other authors, so final grades were averaged over the
four authors. The output list of candidates corresponds to average grades $\geq 2$.

Nonlenses are galaxies clearly identified as nearby spirals, ring galaxies, groups, or other contaminants from their morphology,
or cutouts with artifacts. Candidates listed as grade 1 have faint companions or weakly distorted features suggesting possible
strong lensing signatures, but may also correspond to galaxy satellites or spiral arms. Probable lenses show multiple elongated
sources with similar colors, and orientation and angular separation expected for counter-images, while the available 3-color images
cannot firmly rule out contaminants. Those assigned grades of 3 have similar, although brighter, nonblended and unambiguous
signatures of galaxy-scale strong lenses.

\section{Results and discussion}
\label{sec:results}

\subsection{Final candidates from visual inspection}
\label{ssec:cnncand}

\begin{table*}
\centering
\begin{tabular}{lcccccccccccc}
\hline
\hline \\[-0.3em]
Name & RA & Dec & $p_{\rm CNN}$ & Grade & $g_{\rm Kron}$ & $r_{\rm Kron}$ & $i_{\rm Kron}$ & $g_{\rm aper}$ & $r_{\rm aper}$ & $i_{\rm aper}$ & Redshift & Notes \\[+0.5em]
\hline \\[-0.3em]
  PS1J2226$+$0041 & 22:26:09 & $+$00:41:42 & 0.990 & 3.00 & ...   & 20.59 & 19.37 & 23.34 & 21.9 & 20.63 & 0.6471$^{(*)}$ & (a), (b), (c) \\
  PS1J1821$+$7130 & 18:21:40 & $+$71:30:10 & 1.000 & 3.00 & 18.66 & 17.61 & 17.19 & 20.12 & 19.04 & 18.62 & ... & \\             
  PS1J1647$+$1117 & 16:47:04 & $+$11:17:49 & 0.951 & 3.00 & 19.47 & 18.56 & 18.00 & 20.98 & 19.92 & 19.32 & ... & \\            
  PS1J1559$+$3147 & 15:59:23 & $+$31:47:12 & 1.000 & 3.00 & 18.19 & 17.18 & 16.58 & 19.77 & 18.74 & 18.15 & 0.1489$^{(*)}$ & \\  
  PS1J1508$-$1652 & 15:08:10 & $-$16:52:38 & 1.000 & 3.00 & ...   & 18.95 & 18.39 & 22.42 & 20.65 & 19.88 & ... & \\            
  PS1J1421$-$0536 & 14:21:28 & $-$05:36:51 & 1.000 & 3.00 & 19.57 & 18.59 & 17.97 & 21.22 & 19.99 & 19.29 & ... & \\            
  PS1J1415$+$1112 & 14:15:31 & $+$11:12:08 & 0.993 & 3.00 & 19.54 & 18.30 & 17.65 & 21.05 & 19.81 & 19.16 & 0.3155$^{(*)}$ & \\  
  PS1J1322$-$0501 & 13:22:27 & $-$05:01:34 & 0.995 & 3.00 & ...   & 20.18 & 19.33 & 23.26 & 21.55 & 20.66 & ... & \\            
  PS1J1113$+$0104 & 11:13:57 & $+$01:04:05 & 0.951 & 3.00 & ...   & 20.77 & 19.49 & 23.21 & 21.81 & 20.70 & 0.6402$^{(*)}$ & \\    
  PS1J0919$+$0336 & 09:19:05 & $+$03:36:39 & 0.919 & 3.00 & 20.28 & 19.12 & 18.41 & 22.12 & 20.68 & 19.90 & 0.4440$^{(*)}$ & (c), (l)  \\
  PS1J0353$-$1706 & 03:53:46 & $-$17:06:39 & 0.933 & 3.00 & ...   & 20.35 & 19.50 & 23.47 & 21.72 & 20.63 & ... & (a) \\
  PS1J0324$-$1020 & 03:24:49 & $-$10:20:53 & 0.999 & 3.00 & 20.72 & 19.83 & 18.93 & 22.48 & 21.20 & 20.22 & ... & \\          
  PS1J0211$-$1938 & 02:11:01 & $-$19:38:10 & 0.944 & 3.00 & 20.42 & 19.29 & 18.80 & 22.05 & 20.66 & 19.97 & ... & \\          
  PS1J0141$-$1713 & 01:41:06 & $-$17:13:24 & 1.000 & 3.00 & ...   & 20.08 & 19.19 & 23.30 & 21.54 & 20.62 & 0.56 & (a), (d) \\
  PS1J2348$+$0148 & 23:48:45 & $+$01:48:35 & 0.989 & 2.75 & 22.00 & 20.72 & 19.32 & 23.12 & 21.64 & 20.43 & 0.5902$^{(*)}$ & \\  
  PS1J2336$-$0207 & 23:36:10 & $-$02:07:35 & 1.000 & 2.75 & 20.43 & 19.52 & 18.91 & 21.92 & 20.80 & 20.06 & 0.4942$^{(*)}$ & \\    
  PS1J2248$-$0103 & 22:48:01 & $-$01:03:00 & 1.000 & 2.75 & 19.44 & 18.15 & 17.56 & 20.92 & 19.62 & 18.95 & 0.2772$^{(*)}$ & (e)  \\
  PS1J2247$+$1801 & 22:47:20 & $+$18:01:21 & 1.000 & 2.75 & 19.97 & 18.53 & 17.96 & 21.59 & 20.16 & 19.57 & 0.3427$^{(*)}$ & \\   
  PS1J2233$+$3012 & 22:33:34 & $+$30:12:23 & 0.998 & 2.75 & 19.93 & 18.31 & 17.68 & 21.64 & 20.06 & 19.41 & 0.3605$^{(*)}$ & \\    
  PS1J2202$+$0614 & 22:02:52 & $+$06:14:50 & 1.000 & 2.75 & 18.52 & 17.40 & 16.83 & 19.87 & 18.78 & 18.22 & 0.17 & \\         
  PS1J2200$-$1024 & 22:00:47 & $-$10:24:27 & 0.997 & 2.75 & 20.82 & 19.07 & 18.34 & 22.02 & 20.15 & 19.45 & ... & \\          
  PS1J1926$-$2138 & 19:26:18 & $-$21:38:20 & 1.000 & 2.75 & 16.48 & 15.90 & 15.65 & 17.80 & 16.98 & 16.69 & ... & \\        
  PS1J1749$+$2330 & 17:49:12 & $+$23:30:36 & 0.992 & 2.75 & ...   & 18.65 & 17.86 & 22.06 & 20.37 & 19.76 & ... & \\          
  PS1J1655$+$0406 & 16:55:09 & $+$04:06:13 & 1.000 & 2.75 & 18.30 & 17.77 & 17.32 & 20.03 & 19.33 & 18.86 & ... & \\         
  PS1J1644$+$1121 & 16:44:08 & $+$11:21:30 & 0.998 & 2.75 & 17.85 & 17.28 & 16.87 & 20.03 & 19.17 & 18.60 & 0.14 & \\         
  PS1J1555$+$4151 & 15:55:18 & $+$41:51:39 & 1.000 & 2.75 & 20.96 & 19.93 & 19.02 & 22.25 & 21.32 & 20.35 & 0.5553$^{(*)}$ & (c) \\
  PS1J1553$-$0142 & 15:53:38 & $-$01:42:34 & 0.995 & 2.75 & 21.26 & 19.87 & 18.80 & 22.90 & 21.30 & 20.32 & 0.40 & \\       
  PS1J1445$+$3649 & 14:45:54 & $+$36:49:49 & 0.960 & 2.75 & 19.85 & 18.33 & 17.66 & 21.06 & 19.52 & 18.89 & 0.3551$^{(*)}$ & \\   
  PS1J1439$+$0721 & 14:39:37 & $+$07:21:01 & 0.913 & 2.75 & 20.80 & 19.48 & 18.65 & 22.25 & 20.63 & 19.79 & 0.4795$^{(*)}$ & \\   
  PS1J1422$+$4246 & 14:22:41 & $+$42:46:08 & 0.983 & 2.75 & 21.43 & 20.46 & 19.63 & 22.89 & 21.86 & 20.82 & 0.6047$^{(*)}$ & (e) \\
  PS1J1411$+$2313 & 14:11:05 & $+$23:13:57 & 0.955 & 2.75 & 19.80 & 18.19 & 17.58 & 21.29 & 19.74 & 18.93 & 0.3508$^{(*)}$ & \\   
  PS1J1349$+$0537 & 13:49:15 & $+$05:37:51 & 0.995 & 2.75 & 20.78 & 19.57 & 18.64 & 22.33 & 20.75 & 19.80 & 0.5055$^{(*)}$ & \\   
  PS1J1241$+$2721 & 12:41:37 & $+$27:21:25 & 1.000 & 2.75 & 20.70 & 20.15 & 19.64 & 22.10 & 21.40 & 20.59 & 0.50 & \\      
  PS1J1206$+$2336 & 12:06:51 & $+$23:36:02 & 1.000 & 2.75 & 19.95 & 18.92 & 18.46 & 21.57 & 20.11 & 19.62 & 0.25 & \\   
  PS1J1111$+$4329 & 11:11:55 & $+$43:29:11 & 1.000 & 2.75 & ...   & 20.01 & 19.31 & 23.21 & 21.75 & 20.57 & 0.6370$^{(*)}$ & \\   
  PS1J0921$+$0214 & 09:21:37 & $+$02:14:09 & 0.979 & 2.75 & 19.57 & 18.21 & 17.71 & 21.41 & 20.09 & 19.42 & 0.3191$^{(*)}$ & (f), (l) \\ 
  PS1J0907$+$4233 & 09:07:28 & $+$42:33:02 & 1.000 & 2.75 & 19.67 & 18.75 & 18.05 & 21.32 & 20.40 & 19.51 & 0.4950$^{(*)}$ & \\     
  PS1J0737$+$1914 & 07:37:10 & $+$19:14:36 & 0.989 & 2.75 & 20.03 & 18.90 & 18.31 & 21.69 & 20.25 & 19.54 & 0.3528$^{(*)}$ & \\          
  PS1J0717$+$4624 & 07:17:41 & $+$46:24:31 & 1.000 & 2.75 & 19.61 & 18.02 & ...   & 21.18 & 19.65 & 19.12 & ... & \\         
  PS1J0520$-$2008 & 05:20:03 & $-$20:08:04 & 0.929 & 2.75 & 19.19 & 18.78 & 18.44 & 20.57 & 20.13 & 19.85 & ... & \\[+0.5em]
\hline
\end{tabular}
\caption{
  Final list of galaxy-scale strong lens candidates with lens LRGs from our systematic search in Pan-STARRS. These systems were selected
  as high confidence candidates with CNN scores $>0.9$, and average grades $\geq 2.0$ from visual inspection. Columns are: source name;
  right ascension; declination; output score from the CNN; average of the visual grades from four authors; $g$-, $r-$, and $i$-band Kron
  magnitudes of the lens and source blends from the PS1 catalog; $g$-, $r-$, and $i$-band aperture magnitudes of 1.04\arcsec\ radii covering
  the lens central regions; SDSS photometric redshifts or spectroscopic redshifts marked as $^{(*)}$ where available; previously published
  confirmed or candidate systems (grades A and B or equivalent). References are the following: (a) \citet{jacobs19b}, (b) \citet{diehl17},
  (c) \citet{sonnenfeld18}, (d) \citet{huang19}, (e) \citet{wong18}, (f) \citet{petrillo19a}, (g) \citet{stark13}, (h) \citet{auger09},
  (i) \citet{jacobs19a}, (j) \citet{lemon19}, (k) \citet{wang17}, (l) \citet{jaelani20}, and (m) \citet{schirmer10}. The complete table
  is available in the online version of this paper.}
\label{tab:cand}
\end{table*}

Out of the 1.1 million galaxies the CNN scores 598\,130 with $p_{\rm CNN}=0$, and 105\,760, 12\,382, and 1714 as candidate lenses
with $p_{\rm CNN} > 0.5$, $p_{\rm CNN} > 0.9$, and $p_{\rm CNN} = 1$, respectively. Scores $p_{\rm CNN} > 0.5$ amount to 10\% of sources
ranked by the CNN, and only 0.5\% of the input catalog of 23.1 million galaxies. The human inspection process is necessary to
increase purity of the candidate sample. However, a systematic visual classification of galaxies with $p_{\rm CNN} > 0.5$ would be
unrealistic and we use a higher $p_{\rm CNN}$ threshold which impacts the purity and completeness in a way that is difficult to
quantify. Predictions for mock lenses on the test set (see Fig.~\ref{fig:testing}) and for known lenses (see Fig.~\ref{fig:testnet})
suggest that the majority of good candidates have scores $p_{\rm CNN} > 0.9$ with only a few more in the range $0.5 < p_{\rm CNN} < 0.9$,
and we therefore started inspecting all 12\,382 with $p_{\rm CNN} > 0.9$. As the fraction of visual grades $\geq 2$ quickly drops
with decreasing CNN scores, down to $\lesssim1$\% when extending to $0.8<p_{\rm CNN}<0.9$, we restricted our final classification
to $p_{\rm CNN} > 0.9$.

Our selection results in 321 high-confidence candidates with grades $\geq 2$ (hereafter, grades refer to the average grades from
the visual inspectors). Respective fractions of 36\%, 34\%, and 30\% of these candidates have scores in the intervals
$0.99<p_{\rm CNN} \leq 1.00$, $0.95<p_{\rm CNN} \leq 0.99$ and $0.90<p_{\rm CNN} \leq 0.95$, which demonstrates that the CNN learns
meaningful information and assigns high scores for most of the probable or definite lenses. The rate of grades $\geq 2$ over the
12\,382 galaxies with $p_{\rm CNN}>0.9$ is about 2.5\%. This fraction is slightly lower but comparable to previous CNN lens searches
using deeper, higher quality imaging surveys \citep[KiDS DR4, DES Year 3,][]{petrillo19a,jacobs19b}. This is not surprising as some
of these searches have focused on catalogs of LRGs with robust photometry and are less subject to contamination by low-redshift
spirals. Obtaining equivalent performance suggests that our initial catalog-level classification plays an important role in our
systematic search that balances the suboptimal imaging quality of the PS1 3$\pi$ survey. Finally, 37 additional candidates from
previous CNNs with $p_{\rm CNN} = 1.0$ and grades $\geq 2$ are included (see Sect.~\ref{ssec:perf}), bringing the total number of
resulting lens candidates to 358 from our two-step search.

Examples of good candidates and false positives (grades $\leq 1$) after visual inspection are shown in Fig.~\ref{fig:cnncand}. In
most cases, false positives belong to specific galaxy types frequently misclassified by the network such as nearly face-on spirals
with obvious and extended arms, nearby lenticular galaxies, and bright LRGs with faint unlensed companions. Galaxy groups, mergers
with perturbed morphologies, and cutouts with background artifacts are more rare. Other ambiguous systems listed as false positives
are galaxies with red bulges and faint arms, with color gradients or companions, where all components are blended and mimic lensed
arcs.

\begin{figure*}
\centering
\includegraphics[height=.28\textheight]{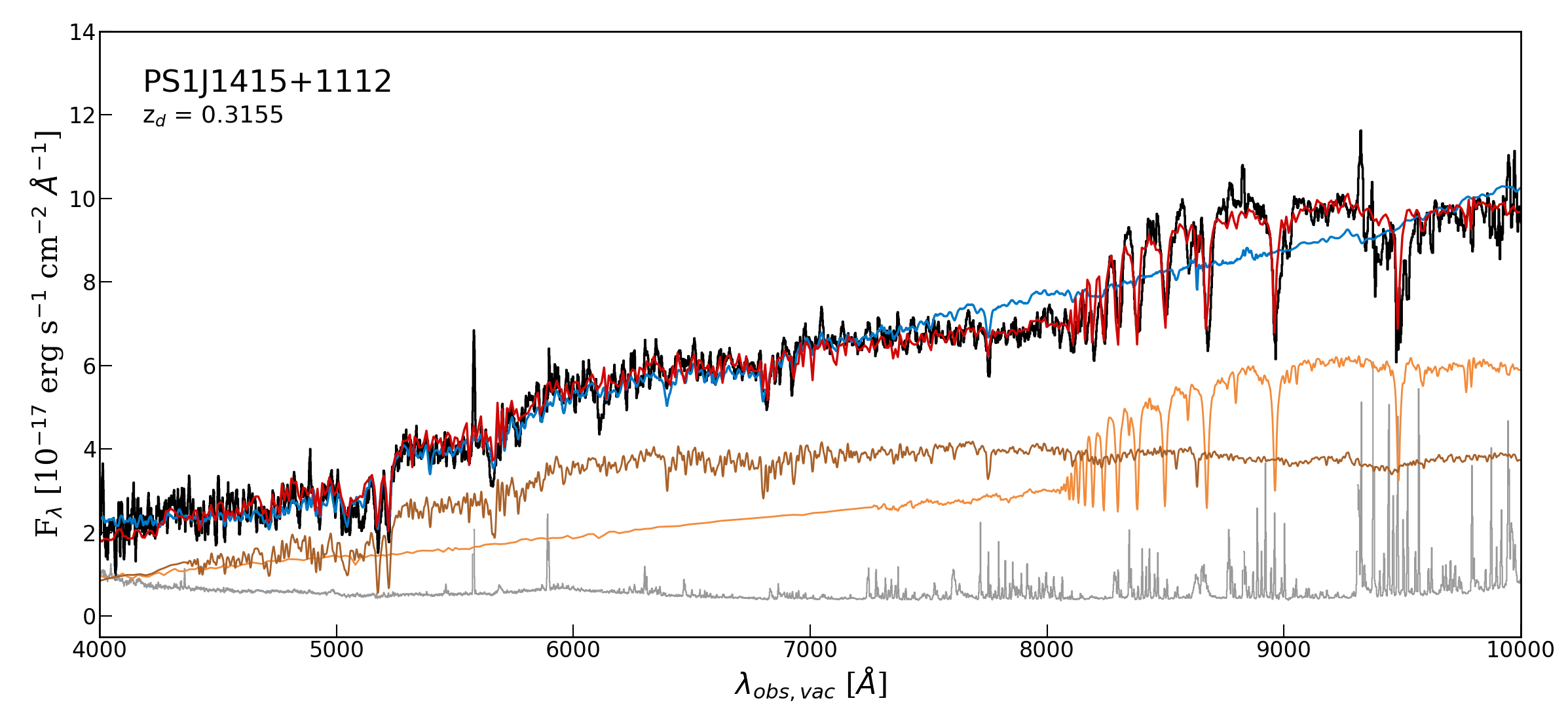}
\caption{
  Pan-STARRS lens candidate with prominent Balmer absorption features from a background galaxy at $z_{\rm s}=1.185$ blended with
  the BOSS spectrum of the lens LRG at $z_{\rm d}=0.3155$. The black and gray lines show the spectrum observed with a fiber of
  2\arcsec\ diameter and the 1$\sigma$ noise level, respectively, and the blue line corresponds to the best-fit LRG template from
  the automated SDSS pipeline at $\lambda < 8000$~\AA\ (see inset), which poorly fits the continuum and Balmer lines at longer
  wavelengths. The red line shows that combining the SED templates of an LRG at $z_{\rm d}=0.3155$ (brown curve) and a recently
  quenched galaxy at $z_{\rm s}=1.185$ (orange curve) correctly fits the spectrum over the entire range (see details in text).}
\label{fig:boss_conf1}
\end{figure*}

The PS1 lens candidates were cross-matched with galaxy-scale strong lenses from previous searches with the status of spectroscopally
confirmed or candidate systems, by building upon and expanding the current MasterLens database. We compiled grade A and B systems (or
equivalent) from a number of recent searches based on optical and near-infrared imaging from DES \citep{diehl17,jacobs19a,jacobs19b},
CFHTLS \citep{more16,jacobs17}, KiDS \citep{petrillo17,petrillo19a,li20}, DESI \citep{huang19}, and HSC \citep{wong18,sonnenfeld18,
  jaelani20,sonnenfeld20}. Since our network may also be sensitive to lensed quasars with colors and configurations similar to our
mock lenses, we also cross-matched with the all-sky database of $\sim$220 confirmed lensed quasars in the literature \citep[][and
  references therein]{lemon19}\footnote{https://www.ast.cam.ac.uk/ioa/research/lensedquasars/}, and with previously identified
lensed quasars from HSC \citep{chan19} and from PS1 \citep{rusu19}. For the candidates not included in those tables, we searched
in the SIMBAD Astronomical Database\footnote{http://simbad.u-strasbg.fr/simbad/sim-fcoo}.

To our knowledge, besides the test lenses from the MasterLens database, 23 of our 358 CNN candidates are already listed in the
literature and corresponding references are listed in Table~\ref{tab:cand}\footnote{In addition, some systems have also been found
  independently and concurrently in DESI Legacy Surveys DR8 by \citet{huang20}: PS1J0717$+$4624, PS1J0324$-$1020, PS1J1749$+$2330,
  and PS1J1903$+$5225 are assigned A or B grades and 16 other candidates are assigned C grades by \citet{huang20}.}. The vast majority
of them are also galaxy-scale strong lens candidates from ground-based multiband imaging searches and lack both spectroscopic and
high-resolution imaging follow-up. The only galaxy-galaxy lens systems confirmed with spectroscopy or space-based imaging are
PS1J0143$+$1607, PS1J0145$-$0455, PS1J2343$-$0030, and PS1J0454$-$0308. Firstly, PS1J0143$+$1607 is CSWA~116, discovered through the
search of blue lensed features near LRGs in SDSS imaging \citep{stark13}. Its properties ($z_{\rm d} = 0.415$, $z_{\rm s} = 1.499$,
$\theta_{\rm E} = 2.7$\arcsec) are very similar to our mock lenses, and the system was assigned $p_{\rm CNN} = 1.0$ and a visual grade
of 2.75. Secondly, PS1J0145$-$0455 is CSWA~103, also presented by \citet{stark13}, and has $z_{\rm d} = 0.633$, $z_{\rm s} = 1.958$, and
$\theta_{\rm E} = 1.9$\arcsec, akin to our mocks. It has $p_{\rm CNN} = 1.0$ and a visual grade of 2.50. Thirdly, PS1J2343$-$0030 is a
SLACS lens with $z_{\rm d} = 0.181$ and $z_{\rm s} = 0.463$ which was discovered and modeled by \citet{auger09} using SDSS spectroscopy
and HST imaging. It has an Einstein radius of 1.5\arcsec, higher than the majority of lenses in the SLACS sample, which explains its
elevated scores from the CNN ($p_{\rm CNN} = 0.917$) and PS1 cutout inspection (2.50). Lastly, PS1J0454$-$0308 ($p_{\rm CNN} = 1.000$ and
grade = 2.75) is a peculiar system thoroughly studied in \citet{schirmer10}, where the main lens is the brightest elliptical galaxy
of a fossil group at $z=0.26$ only 8\arcmin\ from the MS0451$-$0305 cluster at $z=0.54$. The HST F814W frame clearly resolves the
background source into an extended arc and a compact counter image, corresponding to $\theta_{\rm E} = 2.4$\arcsec. Most other
galaxy-galaxy lenses in the literature were missed by our selection due to limited PS1 depth or lens configurations not represented
in our mocks (e.g., higher Einstein radii). Finally, two of these 23 published systems are confirmed lensed quasars. PS1J2350+3654
($p_{\rm CNN} = 1.0$ and grade of 2.25) was discovered in Gaia DR2 \citep{lemon19}, and PS1J1640+1932 ($p_{\rm CNN} = 1.0$ and grade of
2.25) from SDSS \citep{wang17}.

Postage stamps of the 335 newly-discovered and 23 published galaxy-scale lens candidates from our Pan-STARRS CNN search are shown
in Fig.~\ref{fig:cnncand} and in Appendix A, together with $p_{\rm CNN}$ and visual inspection grades. Given that some visual
identifications rely on Legacy imaging, which is sometimes deeper than PS1, we also show Legacy 3-color images of our candidates
located in the Legacy footprint in Appendix A.

\subsection{Ancillary spectroscopy}
\label{ssec:match}

We inspected SDSS BOSS spectra from the 16$^{th}$ data release available for 104 out of 358 lens candidates, in order to characterize
the candidate lens galaxies and to search for spectral signatures of high-redshift background galaxies. This approach was previously
used to select the SLACS sample \citep{bolton08} and relies on spectral features captured within the small, 2\arcsec\ diameter aperture
fibers. Firstly, our examination results in 84 spectra of typical LRGs at intermediate redshift with bright continuum, prominent
4000\AA\ break, deep stellar absorption lines, and non- or very faint [OII]$\lambda\lambda$3727 detections indicating evolved stellar
populations with little residual star formation. Secondly, we obtain seven LRG-like spectra with two or more emission or absorption
lines falling at a different and concordant redshift, higher than the LRG redshift, and indicating the presence of a background galaxy.
Two of them are already published as confirmed strong lens systems: PS1J2343$-$0030 as part of the SLACS sample \citep{auger09}, and
PS1J1640$+$1932 \citep[lensed quasar from][]{wang17}. Thirdly, we find eight LRG-like spectra overlaid with a single bright,
high-redshift emission line consistent with [OII]$\lambda\lambda$3727 from a background star-forming galaxy at $0.95<z<1.50$. Although
the line widths and resolved double-peaked profiles are those expected for the [OII] doublet, other bright emission lines
(H$\beta\lambda$4863, [OIII]$\lambda$4960 and [OIII]$\lambda$5008) are redshifted out of the BOSS spectral window and we consider these
identifications as ambiguous. Lastly, three cases show clear signatures of star-forming galaxies at $z<0.3$, which likely have blue
arms misidentified as lensed arcs, one QSO at $z=0.5890$, and one star. These five systems are considered as false positives.

Finding a vast majority of LRGs\footnote{Although we can not completely rule out that in few cases, these are actually the central
  bulges of late-type galaxies covered by BOSS.} and only 3/104 star-forming galaxies at lower redshift demonstrates the validity of
our method. This result shows that our CNN and visual inspection method predominantly selects the targeted population of galaxy-scale
strong lens candidates with lens LRGs, and efficiently distinguishes strong lensing features from usual interlopers (e.g., spiral
arms, tidal features, blue rings). Moreover, the five false positives (PS1J2249$+$3228, PS1J1551$+$2156, PS1J1452$+$1047,
PS1J0834$+$1443, and the QSO PS1J1156$+$5032) are all listed as lower confidence candidates with grades $\leq 2.25$.

\begin{figure*}
\centering
\includegraphics[height=.28\textheight]{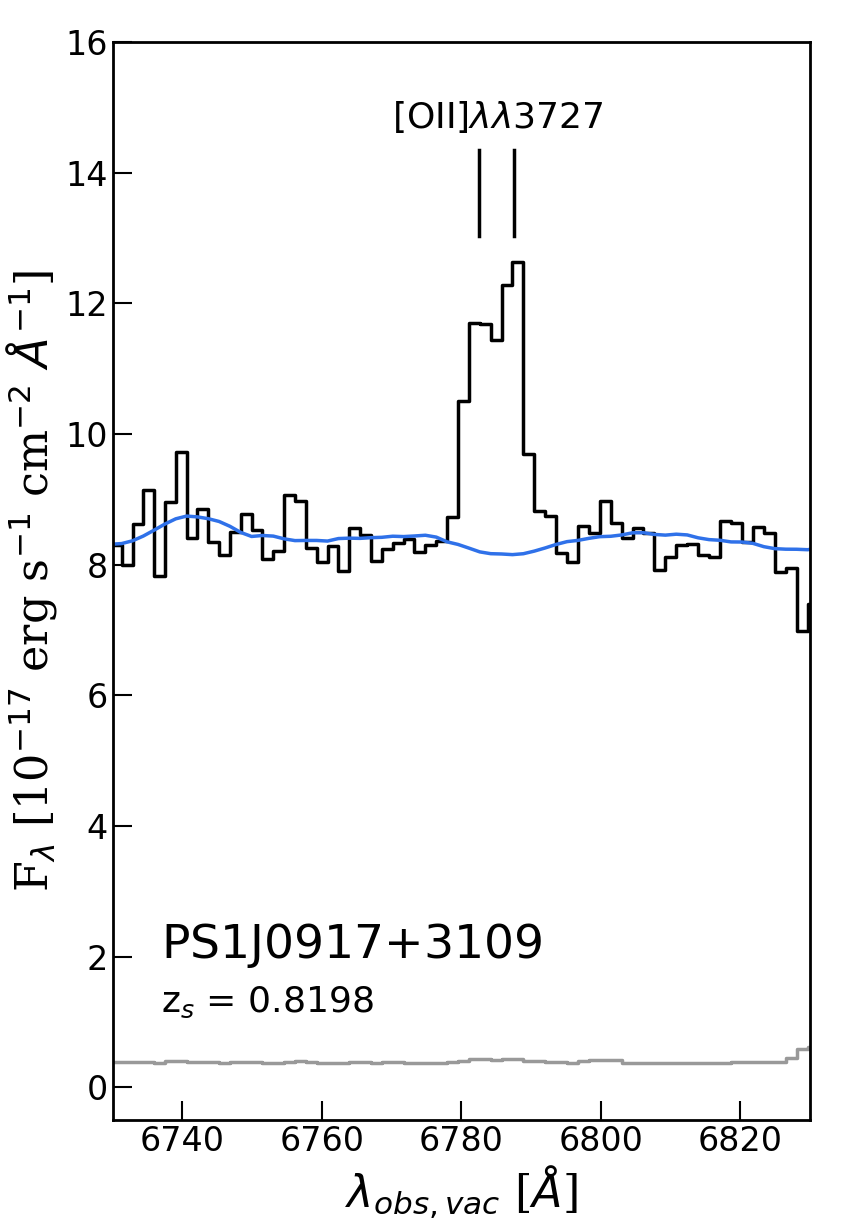}
\includegraphics[height=.28\textheight]{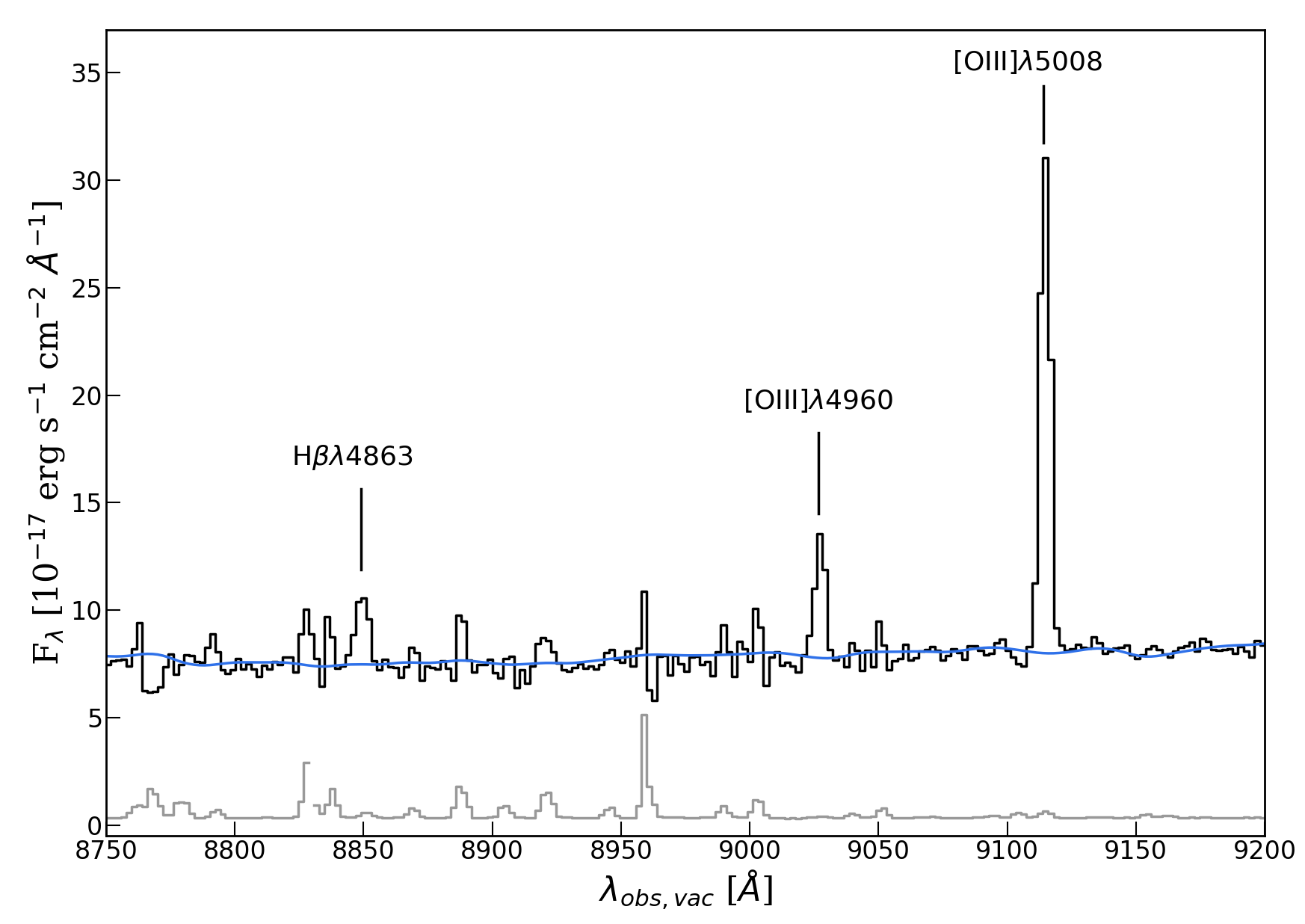}
\caption{
  Example of a lens candidate with multiple emission lines from a high-redshift background galaxy overlaid on the SDSS BOSS spectrum
  of the foreground LRG. The black, gray and blue lines show the observed spectrum, the 1$\sigma$ noise level, and the best-fit SDSS
  template for the LRG, respectively. The spectrum is zoomed on the spectral features associated with the background line-emitter at
  $z = 0.8198$ rather than with the LRG at $z = 0.2386$.}
\label{fig:boss_conf2}
\end{figure*}

The rarity of spectroscopic signatures from lensed galaxies is not surprising because the 2\arcsec\ BOSS fibers enclose little
lensed source emission in the $\theta_{\rm E} \gtrsim 1.5$\arcsec\ systems targeted by our search. Background line flux can only be
detected in few favorable cases, like in the presence of counter-images closer to the lens center. In addition, emission lines in
most lensed galaxies might be too faint for SDSS spectroscopy, and the observable spectral band of BOSS excludes H$\beta$4863,
[OIII]4960 and [OIII]5008 for $z>0.9$, which rules out multiple line detections for the majority of sources expected at $z>1$.

The only Pan-STARRS candidate robustly confirmed as a strong lens system is PS1J1415$+$1112. Pan-STARRS imaging spatially resolves
the system into a quadruply-imaged background source forming a typical fold configuration, and with redder color than the lens LRG
(Fig.~\ref{fig:cnncand}). The best-fit model from the BOSS pipeline primarily identifies spectral features from the foreground LRG
at $z_{\rm d}=0.3155$, but poorly fits the continuum and remarkably prominent Balmer absorption lines at $\lambda_{\rm obs} > 8000$~\AA\
(Fig.~\ref{fig:boss_conf1}). These features are associated with the background galaxy at $z_{\rm s}=1.185$. Further inspection of
the BOSS spectrum shows that the strong Balmer absorption lines, together with the weak 4000~\AA\ break indicative of relatively
young stellar populations, and the lack of nebular emission lines at $z_{\rm s}=1.185$ suggesting little on-going star formation,
favor the scenario of a recently quenched galaxy. To verify this hypothesis, we used the BC03 stellar population synthesis models
\citep{bruzual03} to combine the SED template of an elliptical galaxy at $z_{\rm d}=0.3155$ with 7~Gyr old stellar populations
and solar metallicity with various templates of post-starbursts at $z_{\rm s}=1.185$. After varying the stellar age, metallicity, and
dust attenuation, we found that rescaling the SED of a young galaxy with 25~Myr old stellar populations, no on-going star formation,
and $Z=0.2$~Z$_{\odot}$ correctly fits the overall wavelength range in Fig.~\ref{fig:boss_conf1}. Together with the red colors of
multiple images this adds further evidence that the lensed galaxy has indeed ceased star formation recently.

The last four CNN candidates showing multiple emission lines from a background, possibly lensed galaxy do not exhibit such clear
configurations in Pan-STARRS and Legacy $gri$ images. Their BOSS spectra reveal simultaneous detections of [OII]$\lambda\lambda$3727,
H$\beta\lambda$4863, [OIII]$\lambda$4960, and [OIII]$\lambda$5008 on top of the prominent stellar continuum from the foreground
LRG (see Fig.~\ref{fig:boss_conf2}). The first three are PS1J2345$-$0209 with $z_{\rm LRG}=0.2940$ and $z_{\rm s}=0.6581$, PS1J1134$+$1712
with $z_{\rm LRG}=0.3752$ and $z_{\rm s}=0.8121$, and PS1J0917$+$3109 with $z_{\rm LRG}=0.2386$ and $z_{\rm s}=0.8198$. These candidates
were listed as probable lenses from our grades and, although $z_{\rm LRG}$, $z_{\rm s}$, and $\sigma^*_{\rm LRG}$ suggest plausible
Einstein radii $\sim 1$\arcsec\ for singular isothermal sphere profiles, high-resolution follow-up imaging is needed to ascertain their
configuration. These data will determine whether the background, spectroscopically-confirmed galaxies are those visually identified
with Pan-STARRS and indeed multiply-imaged, or whether they are out of the strong lensing regime. Lastly, PS1J1724$+$3146 comprises
a foreground LRG at $z_{\rm LRG}=0.2097$ and a background source at $z_{\rm s}=0.3461$, likely too close to the lens to enter the strong
lensing regime.

\subsection{Properties of candidates}
\label{ssec:prop}

Our visual inspection stage implies that resulting lens candidates have morphologies and configurations easily detectable by the
human eye. Figure~\ref{fig:cnncand} confirms that higher grades are given to systems with extended arcs, clear counter images and
often compact lens light profiles. Postage stamps make it clear that our most reliable systems have relatively brighter and distorted
arcs (as plausible strongly lensed background galaxies) in $gri$ bands which will facilitate their future follow-up. Interestingly,
besides the numerous blue arcs (e.g., PS1J1647$+$1117, PS1J1508$-$1652), several candidates have lensed sources redder than the
foreground LRGs (e.g., PS1J1559$-$3147, PS1J1445$+$3649) and others have compact, nearly point-like morphologies suggestive of
lensed quasars (e.g., PS1J1926$-$2138). Since the presence of possible counter images is a prerequisite for lens identifications,
our candidate set is presumably biased toward higher fractions of quads than doubles, at least for grades $\sim 3$. Other biases
on the relative positions of lenses and sources might also arise. For instance, although our tests suggest that systems with
$\theta_{\rm E}/R_{\rm eff}<1$ are correctly classified by the CNN, with only minor differences in $p_{\rm CNN}$ with respect to
$\theta_{\rm E}/R_{\rm eff}>1$ (Fig.~\ref{fig:testprop}), candidates and known lenses recovered by visual inspection should be
relatively less blended and lie on the high-end of the $\theta_{\rm E}/R_{\rm eff}$ distribution. Quantifying this effect is
nevertheless an arduous task due to uncertainties in reliably measuring $R_{\rm eff}$ from Pan-STARRS imaging. As previously
mentioned, systems with $\theta_{\rm E}$ differing from simulated lenses in our training set are likely all discarded.

Our 358 candidates have lens redshifts in the range $z_{\rm d} \sim 0.15$--0.65. The distribution does not change much between
candidates with confirmed lens $z_{\rm spec}$ from SDSS, and those with more uncertain SDSS $z_{\rm phot}$ due to lens and source blending.
It closely matches the lens redshift ranges in other samples selected through multiband, ground-based imaging such as CASSOWARY
\citep{stark13} and SL2S \citep{sonnenfeld13}. Interestingly, Pan-STARRS candidates with SDSS stellar velocity dispersion have
$\sigma^*_{\rm mean} \sim 275$~km~s$^{-1}$. This is significantly lower than $\sigma^*_{\rm mean} \sim 310$~km~s$^{-1}$ in the set of
LRGs for simulations (Sect.~\ref{ssec:simu}), indicating that lens LRGs in our candidate set are less massive on average.

To understand implications of such differences in $\sigma^*$ we followed \citet{petrillo17} and, for each lens candidate with
$z_{\rm d}$ and $\sigma^*$ available from BOSS, we calculated Einstein radii for a singular isothermal sphere profile and different
source redshifts spanning the most plausible $z_{\rm s}=0.5$--3.0 range. We assumed $\sigma_{\rm SIS} \sim \sigma^*$. In the vast
majority of cases (86\%), Einstein radii predicted from the lens galaxy dynamics match the approximate
$\theta_{\rm E} \sim 1$--2\arcsec\ of PS1 candidates from a basic inspection of 3-color images in Fig.~\ref{fig:cnncand}. This
range is also expected from construction of the training set and our CNN selection function. Predicted and observed image
configurations usually match for $z_{\rm s}>1$. This qualitative test argues in favor of plausible lens configurations and few
contaminants despite the moderate $\sigma^*$ from BOSS.

In some cases, the large Einstein radii $\gtrsim 1.5$\arcsec\ are likely to be partly attributed to a contribution from the environment
of the primary lens galaxy, probably from a smooth dark-matter halo associated with an extended group- or cluster-scale structure
\citep{oguri06}. As a matter of fact, obtaining a few strong lens candidates with $\theta_{\rm E}$ in the same range as our lens
simulations but with lower SDSS stellar velocity dispersions proves that, if they are genuine strong lenses, they host moderately
massive lens LRGs with significant external shear and magnification contributions from the environment. We verified that some of
our Pan-STARRS candidates are indeed matching the position of the central brightest cluster galaxy of SDSS clusters \citep[e.g.,
  PS1J0907$+$4233, PS1J2153$+$1154, PS1J1417$+$2120, PS1J1210$+$2843, PS1J1730$+$3405,][]{szabo11}, and of candidate galaxy clusters
from a red sequence search in DES imaging \citep[e.g., PS1J0919$+$0336, PS1J2233$+$3012, PS1J1210$+$2843, PS1J1414$+$6447,][]{rykoff16}.
This environmental effect is well illustrated by the case of PS1J0454$-$0308 characterized in \citet{schirmer10}. We also found
that PS1J2210$+$0620 and its bright red arc of $\sim$8\arcsec\ extension lie behind the core of Abell 2422 galaxy cluster
\citep{abell89}. Lastly, PS1J1142$+$5830 is located in the field of the rich, X-ray luminous galaxy cluster MACSJ1142.4$+$5831 at
$z = 0.325$, also known as Abell 1351, which has a bimodal velocity distribution suggesting an on-going merger \citep{barrena14}.
The SDSS photo-z of PS1J1142$+$5830 is consistent with the cluster redshift but our candidate is offset by 1.3\arcmin\ from the
brightest cluster galaxy.

The Pan-STARRS $gri$ photometry of our 358 candidates is comparable to that of our lens simulations. Observed $(g-i)$ colors of
the deflector central regions within 2\arcsec\ diameter apertures range between 1.3 and 2.7~mag and broadly follow the color
distribution of mocks and $p_{\rm cat} > 0.5$ galaxies in Fig.~\ref{fig:catsearch}. Our best candidates are therefore representative
of the global sample and visual inspection does not introduce significant biases in terms of lens color. Kron $(g-i)$ colors
from blended deflector and source emissions are $\sim$0.3 mag lower because lensed arcs are predominantly bluer than the lens
LRGs. Given uncertainties on the lensed source colors, quantitative analyses would require subtracting the lens LRG light
profile, a hazardous task with Pan-STARRS imaging. However, it is worth noting that a majority of $p_{\rm CNN} > 0.9$ systems with
blue arcs closely mimicking extended arms of star-forming spirals were conservatively discarded by visual inspection, while
systems with similar configurations but redder arcs where considered less ambiguous and assigned higher grades. Consequently,
the output set is likely biased toward a higher fraction of red lensed arcs compared to all genuine lens systems assigned
$p_{\rm CNN} > 0.9$.

We did not find significant correlations between average grades and photometric properties. Although our high and lower-confidence
candidates have comparable lens brightness, those with higher grades have slightly redder lenses by $\sim$0.15~mag in $(g-i)$ on
average. Mean lens redshifts and velocity dispersions are also increasing in the top half of the candidate set, by 0.05 and
50~km~s$^{-1}$, respectively.

\subsection{Discussion}
\label{ssec:discu}

Our study demonstrates that a two-step approach combining catalog-level and image-level classifications, as already applied to
lensed quasars searches \citep{agnello15}, is efficient in selecting galaxy-scale strong lenses from very wide-field surveys
without imposing restrictive cuts on the input galaxy catalog. Using three bands and multiaperture photometry, our catalog-level
neural network conveniently reduces the sample size for CNN classification to the order of one million sources which, in the case of
Pan-STARRS, allows us to download all $gri$ cutouts within a manageable amount of time of about one week. This first stage will be
particularly beneficial to overcome data volume limitations in the new era of deep, large scale surveys such as LSST. At the same
time, the low false negative rate and good performance on known lens systems of our catalog-level network applied to Pan-STARRS
photometry shows that this stage discards very few potentially interesting systems. After that, we obtain encouraging results
from our CNN combined with visual inspection. The compilation of 330 newly-discovered lens candidates (after removing the five
false-positives identified from SDSS BOSS spectra) and confirmation of 23 lenses already selected from other, more suitable
surveys (e.g., DES, KiDS) demonstrates the efficiency of our approach regardless of the limited depth and seeing in Pan-STARRS.

The strategy presented in this paper is expected to be easily applicable to the deeper LSST $gri$ stacks in order to identify the
large set of new, wide-separation galaxy-scale lenses from this survey. The simulations of \citet{collett15} suggest that the LSST
final stacks in $gri$ bands will yield about 39\,000 detectable lens systems, with $\theta_{\rm E}$ greater than the seeing FWHM and
large enough to spatially-resolve multiple images, and with total S/N of lensed arcs $>20$ in at least one band. Under these
assumptions, future LSST lenses will mainly cover $\theta_{\rm E}=1$--3\arcsec, a similar range as our Pan-STARRS candidates, and
their systematic identification will be crucial for several science cases including the search for strongly lensed SNe with
spatially-resolved multiple images for early-phase SN spectroscopy and cosmography \citep{suyu20}. In order to achieve this, we
can customize our simulation pipeline of realistic Pan-STARRS mock lenses to produce LSST mocks and quickly assemble a training
set, by using $gri$ cutouts of the relevant LRG lenses from LSST stacks rather than Pan-STARRS. Extending the search to broader
populations of lens galaxies might also become feasible, in light of the $\sim$4~mag deeper $gri$ LSST stacks.

Tests on the CNN performance presented in this paper are likely valid essentially for Pan-STARRS seeing and depth, while better
imaging quality should enable identifications of broader, more complex image patterns and would likely benefit from updates on the
neural network. The main challenge is the minimization of false positives which, given the rarity of strong lenses per deg$^2$, will
largely dominate the sample of CNN candidates even with FPR down to 1\% or less. Potential avenues for improvements include adding
other convolutional layers or boosting the learning process with residual networks \citep{he15}, as the latter might give
better performance for image features not represented in the training data \citep[for LSST simulations, see][]{lanusse18}.
A possibility would be to use CNNs optimized for image outlier detection \citep[e.g.,][]{margalef20}, either to identify strong
lenses as the outlier class, or simply to exclude cutouts with partial coverage, background artifacts or other anomalies before
classification. In addition, although \citet{teimoorinia20} found little difference with single-band HST images, implementing
additional classes for the usual interlopers (spirals, ring galaxies, ...) might help distinguish signatures of strong lenses in
$gri$ bands. Our Pan-STARRS lens search demonstrates that ultimately, most progress should come from the choice of positive and
negative examples in the training set (see Sect.~\ref{ssec:perf}). Highly-realistic lens simulations are one of the main ingredients
\citep[see][]{lanusse18}. For instance, as multiple images are strongly blended with lenses in about half of the mocks produced in
Sect.~\ref{ssec:simu}, one could only include those visually classified as definite lenses from their bright arcs and unambiguous
configurations. Fine-tuning the selection of positive examples in this way could help the network assign high scores exclusively
to robust candidates that are worth including for follow-up campaigns.

Reducing the false positive rates and making sure that only high-confidence candidates get $p_{\rm CNN} \sim 1$ will be crucial for
future lens searches in the LSST era. However, the most efficient neural networks should provide orders of magnitudes more lens
candidates than Pan-STARRS and their systematic visual inspection might become unrealistic. Assembling complete strong lens samples
will therefore require one to abandon this nonautomated, human-classification stage and to search for alternatives. For instance, 
calibrating the CNN scores as probabilities \citep[e.g.,][]{guo17} could help quantify purity and completeness, and automatically
select a subset of candidates for high-resolution imaging or spectroscopic follow-up.

\section{Conclusion}
\label{sec:conclu}

In this paper, we presented a systematic search for wide-separation, galaxy-scale strong lenses in Pan-STARRS using machine learning
classification of $gri$ images from the 3$\pi$ survey on the entire Northern sky. We focused our search on massive LRGs acting as
strong lenses and producing Einstein radii $\geq 1.5$\arcsec, and we simulated a set of highly-realistic mocks by painting lensed
arcs on Pan-STARRS image cutouts of LRGs with known redshift and velocity dispersion from SDSS. This strategy ensures that mocks
include background artifacts and field galaxies, and that the network becomes invariant under the small scale exposure and FWHM
variations on the survey stacks.

We computed the $gri$ aperture photometry of mocks to preselect a conservative catalog of 23.1 million sources, and followed a
two-step approach for classification: (1) a catalog-based neural network on the source photometry, (2) a CNN trained on $gri$ image
cutouts. The catalog-level network assigned $p_{\rm cat}>0.5$ for 1\,050\,207 galaxies while excluding little known lenses from the
MasterLens catalog. The image-level network then yielded sets of 105\,760 and 12\,382 candidates with scores $p_{\rm CNN}>0.5$ and
$>0.9$, respectively. We visually inspected those with $p_{\rm CNN}>0.9$ and combined with previous CNNs to assemble a final set of
330 high-quality newly-discovered candidates with average visual grades $\geq 2$. Publicly available BOSS spectroscopy of the lens
candidates' central regions proves that the vast majority are indeed LRGs at $z \sim 0.1$--0.7, and five newly-discovered candidates
show robust signatures of blended, high-redshift background sources. Pan-STARRS clearly resolves one of them as a quadruply-imaged
red galaxy at $z_{\rm s}=1.185$ (likely recently quenched), behind a lens LRG at $z_{\rm d}=0.3155$. 

This new set of bright lens candidates is particularly valuable for future lensed SNe searches. Strong lenses with such image
separations are indeed likely to produce long time delays of a few days to weeks which alleviates difficulties affecting
$\theta_{\rm E} < 1$\arcsec\ lensed SNe such as iPTF16geu and allows one to measure accurate, microlensing-free time delays for
cosmography. Such time delays are also well-suited to trigger timely imaging and spectroscopic follow-up of a SN reappearance to
characterize the SN's early-phase behavior within a few days after explosion. In the meantime, validating these Pan-STARRS lens
candidates and deriving strong lensing models will require high-resolution imaging and spectroscopic follow-up.

Our CNN exhibits good performance on known lenses detected in Pan-STARRS, correctly classifying 14/16 systems as lenses and
assigning scores $p_{\rm CNN} > 0.9$ to 7/16. We found that CNN predictions strongly depend on the construction of the training set
but little on the network architecture. Our search also recovered 23 confirmed or candidate strong lenses in the literature out
of the MasterLens catalog. In the near future, the release of deep stacks with best seeing conditions in smaller fields as part
of Pan-STARRS DR3 will give the opportunity to extend our search to optimal PS1 data quality. In addition, we expect that the
efficient and automated two-step classification method presented in this paper will be applicable to the deep $gri$ image stacks
from LSST with only minor adjustments.

\section*{Acknowledgements}

We would like to thank the anonymous referee for thoroughly reading the manuscript and providing useful suggestions. We wish to
thank Matt Auger and Alessandro Sonnenfeld for sharing python scripts for spectrum visualization. We also thank B.~Cl\'ement,
F.~Courbin, S.~Huber, C.-H.~Lee, D.~Sluse and A.~Y{\i}ld{\i}r{\i}m for useful discussions and feedback about this work.

RC, SS and SHS thank the Max Planck Society for support through the Max Planck Research Group for SHS. This project has
received funding from the European Research Council (ERC) under the European Union's Horizon 2020 research and innovation
program (LENSNOVA: grant agreement No 771776). This works has also been in part supported by the Swiss National Science
Foundation (SNSF) and by the European Research Council (ERC) under the European Union's Horizon 2020 research and innovation
program (COSMICLENS: grant agreement No 787886).

The Pan-STARRS1 Surveys (PS1) and the PS1 public science archive have been made possible through contributions by
the Institute for Astronomy, the University of Hawaii, the Pan-STARRS Project Office, the Max-Planck Society and its
participating institutes, the Max Planck Institute for Astronomy, Heidelberg and the Max Planck Institute for
Extraterrestrial Physics, Garching, The Johns Hopkins University, Durham University, the University of Edinburgh, the
Queen's University Belfast, the Harvard-Smithsonian Center for Astrophysics, the Las Cumbres Observatory Global Telescope
Network Incorporated, the National Central University of Taiwan, the Space Telescope Science Institute, the National
Aeronautics and Space Administration under Grant No. NNX08AR22G issued through the Planetary Science Division of the
NASA Science Mission Directorate, the National Science Foundation Grant No. AST-1238877, the University of Maryland,
Eotvos Lorand University (ELTE), the Los Alamos National Laboratory, and the Gordon and Betty Moore Foundation.

Based in part on data collected at the Subaru Telescope and retrieved from the HSC data archive system, which is
operated by Subaru Telescope and Astronomy Data Center at National Astronomical Observatory of Japan.
The Hyper Suprime-Cam (HSC) collaboration includes the astronomical communities of Japan and Taiwan, and Princeton
University. The HSC instrumentation and software were developed by the National Astronomical Observatory of Japan
(NAOJ), the Kavli Institute for the Physics and Mathematics of the Universe (Kavli IPMU), the University of Tokyo,
the High Energy Accelerator Research Organization (KEK), the Academia Sinica Institute for Astronomy and Astrophysics
in Taiwan (ASIAA), and Princeton University. Funding was contributed by the FIRST program from Japanese Cabinet Office,
the Ministry of Education, Culture, Sports, Science and Technology (MEXT), the Japan Society for the Promotion of
Science (JSPS), Japan Science and Technology Agency (JST), the Toray Science Foundation, NAOJ, Kavli IPMU, KEK, ASIAA,
and Princeton University. 

This paper makes use of software developed for the Rubin Observatory Legacy Survey of Space and Time. We thank the LSST
Project for making their code available as free software at http://dm.lsst.org.

\bibliography{lenssearch}

\begin{appendix}
  
\section{Complete list of candidates}

\begin{figure*}
\centering
\includegraphics[width=.965\textwidth]{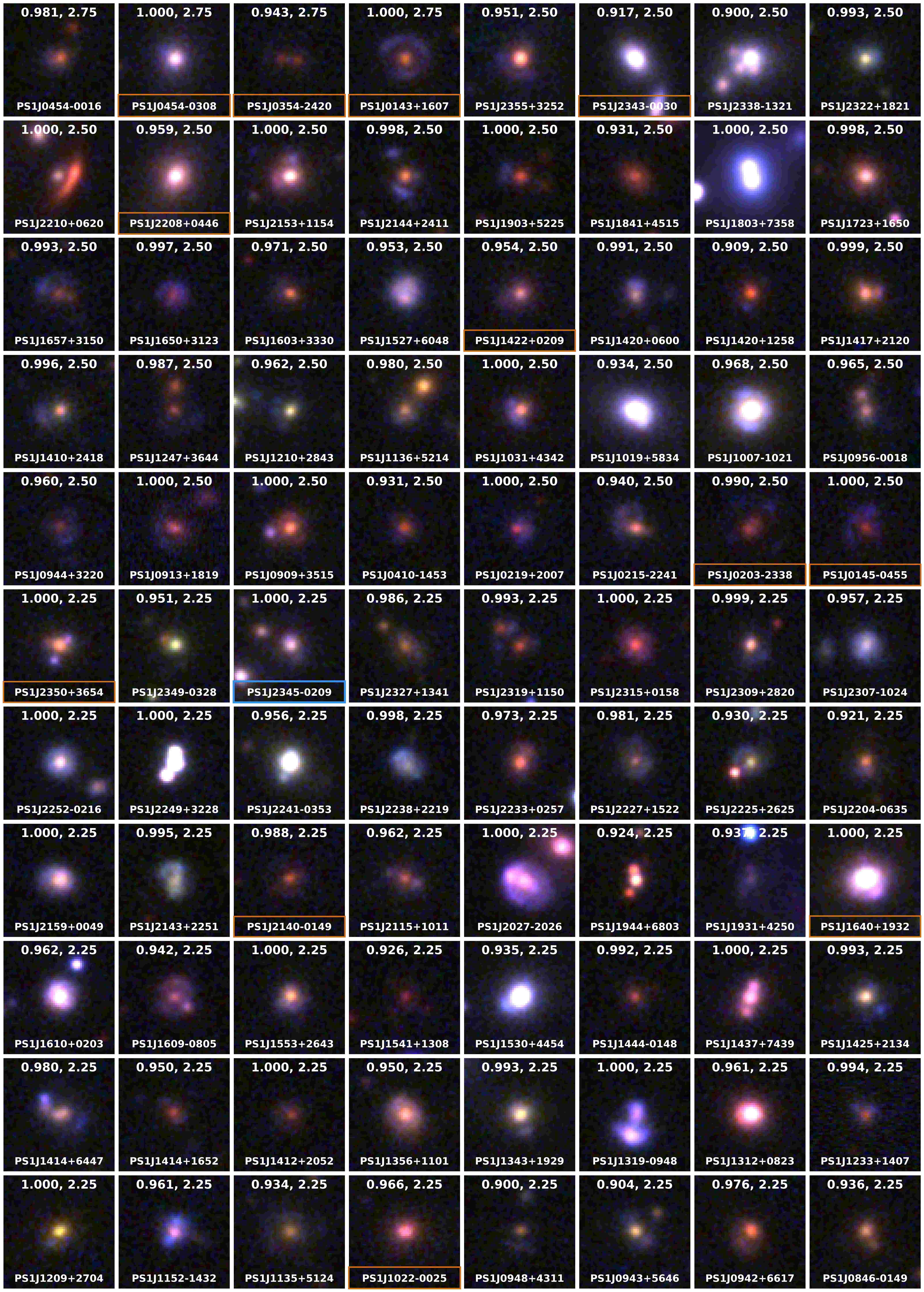}
\caption{Pan-STARRS 3-color $gri$ images of candidates with grades $\geq 2$ from visual inspection of CNN scores $p_{\rm CNN} > 0.9$.
  See Fig.~\ref{fig:cnncand} caption.}
\end{figure*}

\begin{figure*}
\centering
\includegraphics[width=.965\textwidth]{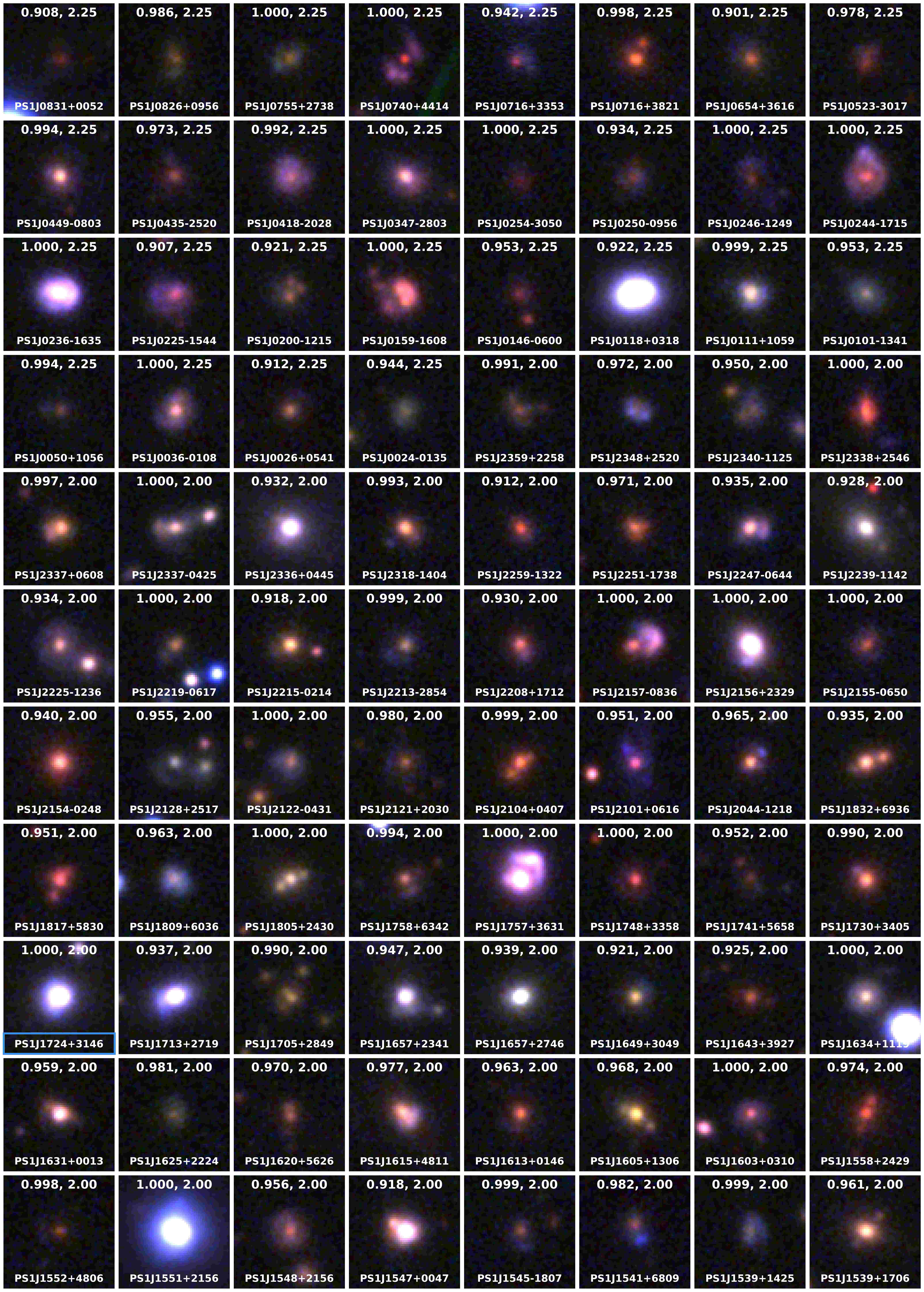}
\caption{continued.}
\end{figure*}

\begin{figure*}
\centering
\includegraphics[width=.965\textwidth]{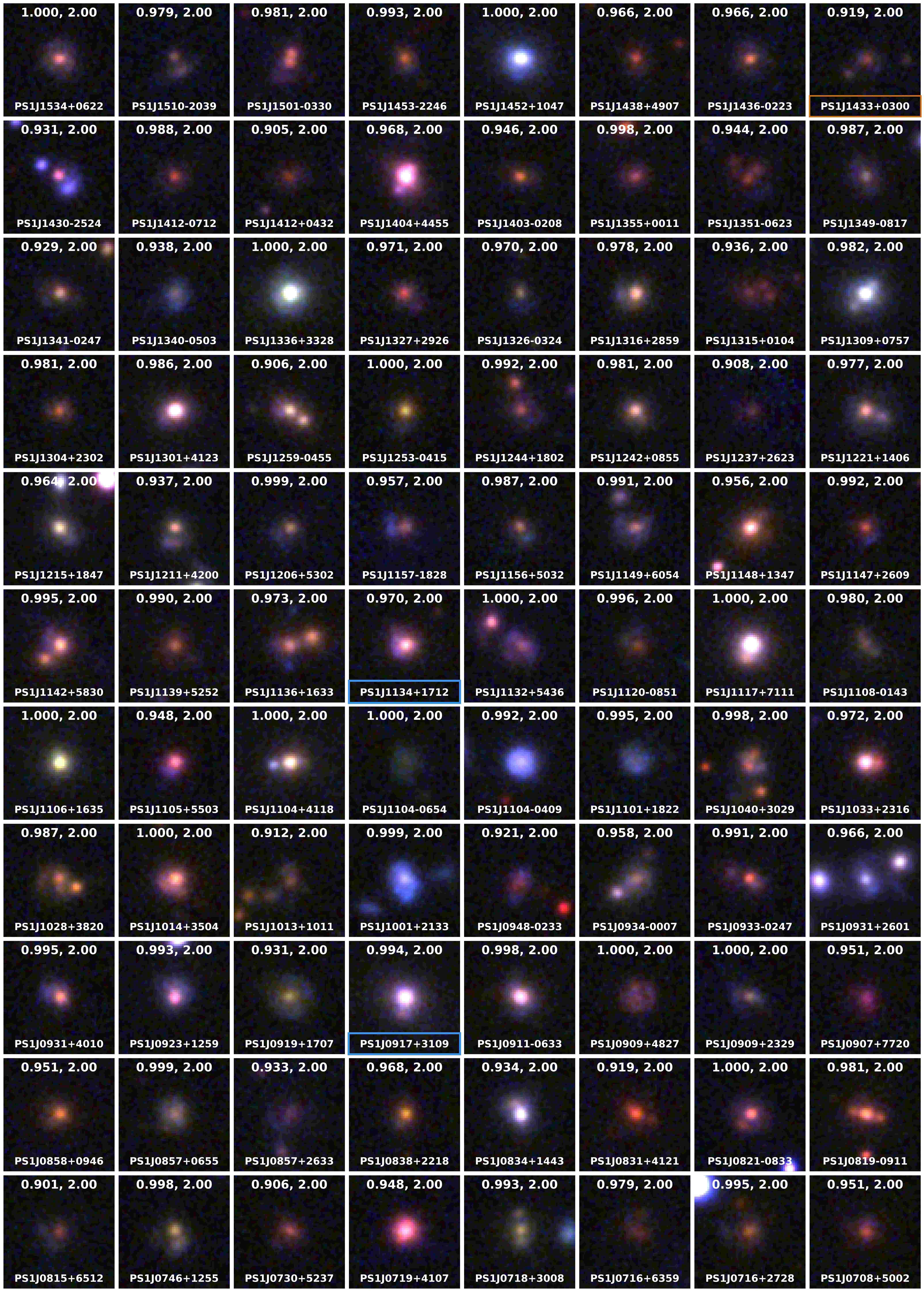}
\caption{continued.}
\end{figure*}

\begin{figure*}
\centering
\includegraphics[width=.965\textwidth]{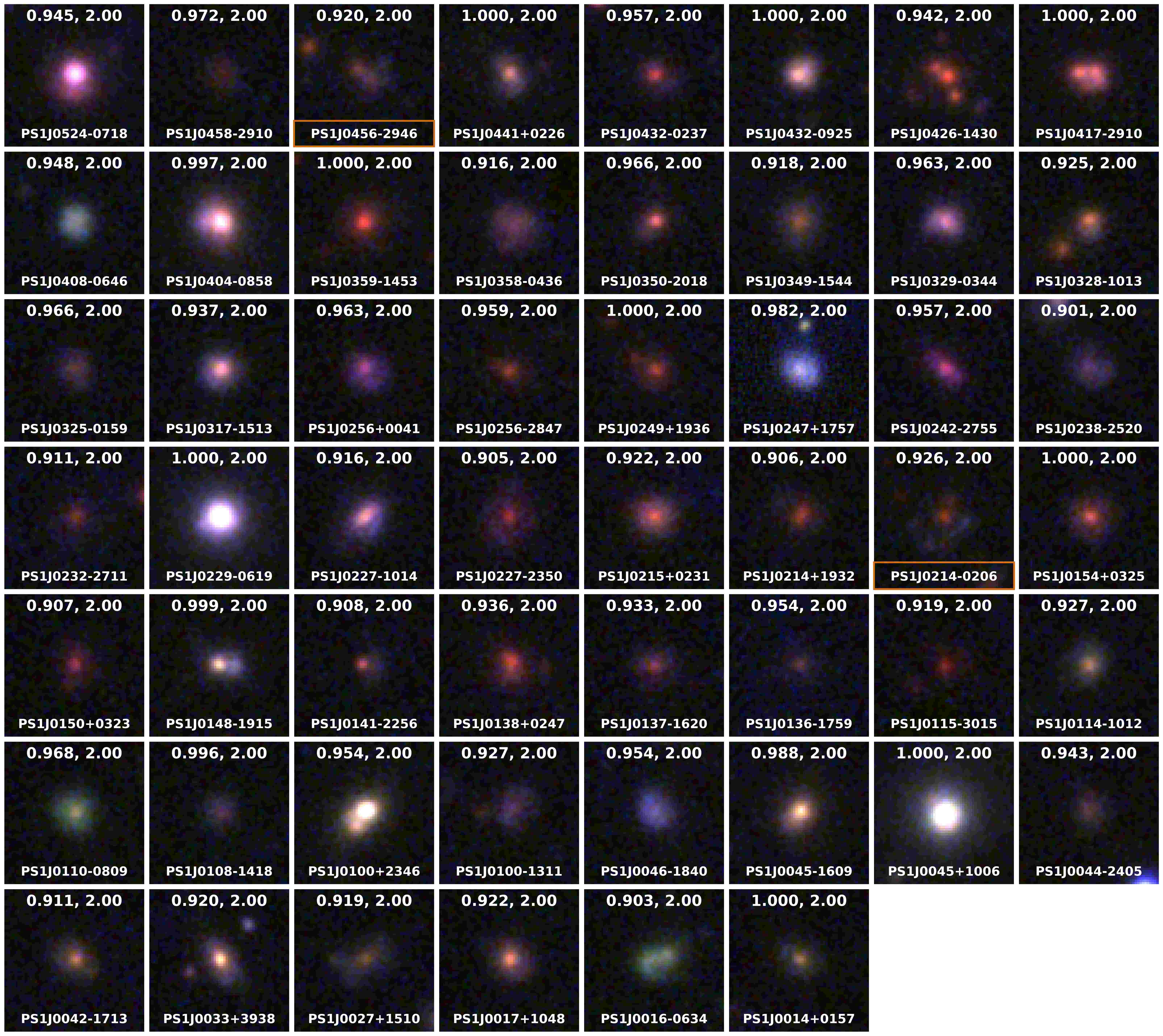}
\caption{continued.}
\end{figure*}

\begin{figure*}
\centering
\includegraphics[width=.965\textwidth]{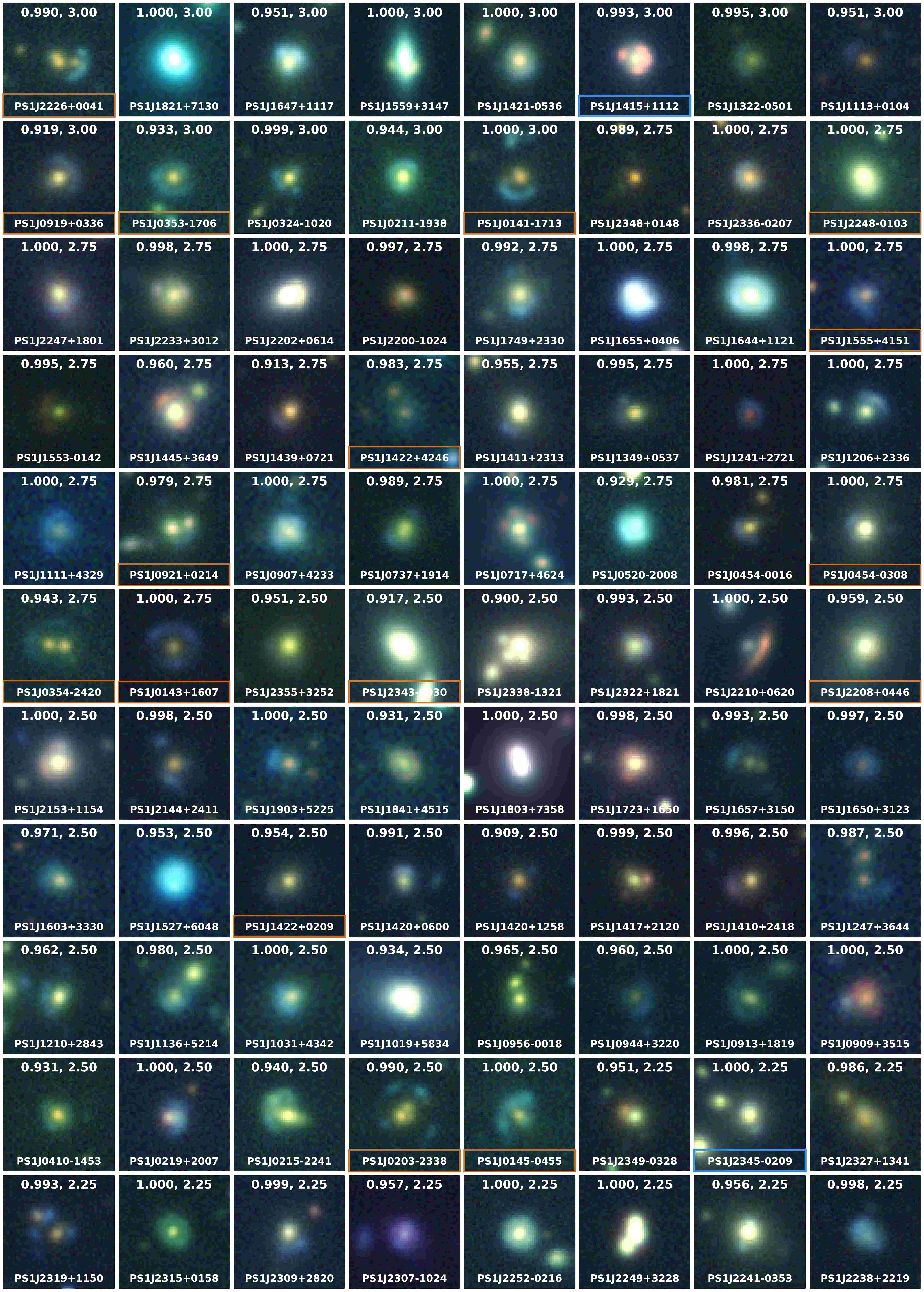}
\caption{Legacy 3-color images of candidates covered in DR8.}
\end{figure*}

\begin{figure*}
\centering
\includegraphics[width=.965\textwidth]{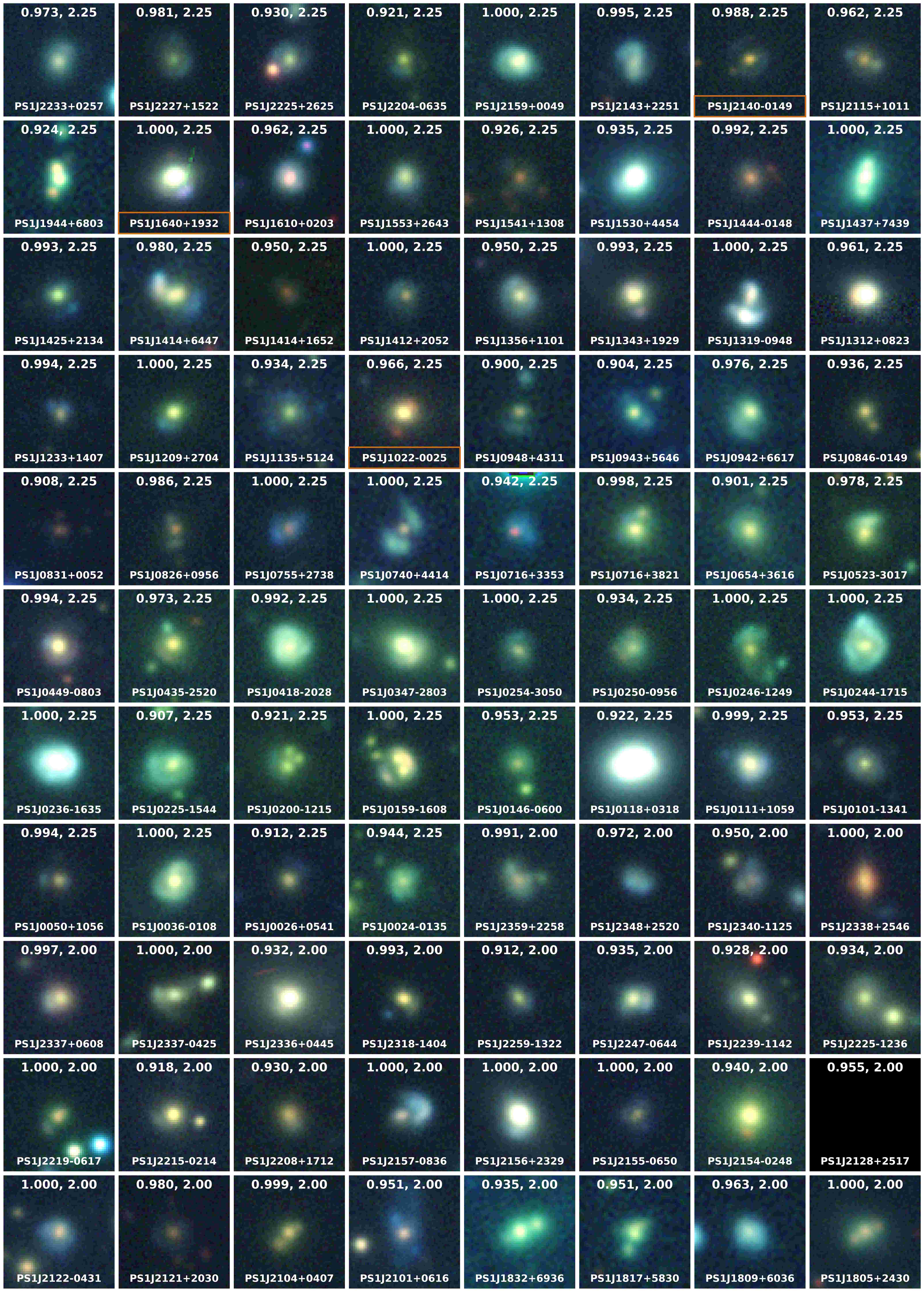}
\caption{continued.}
\end{figure*}

\begin{figure*}
\centering
\includegraphics[width=.965\textwidth]{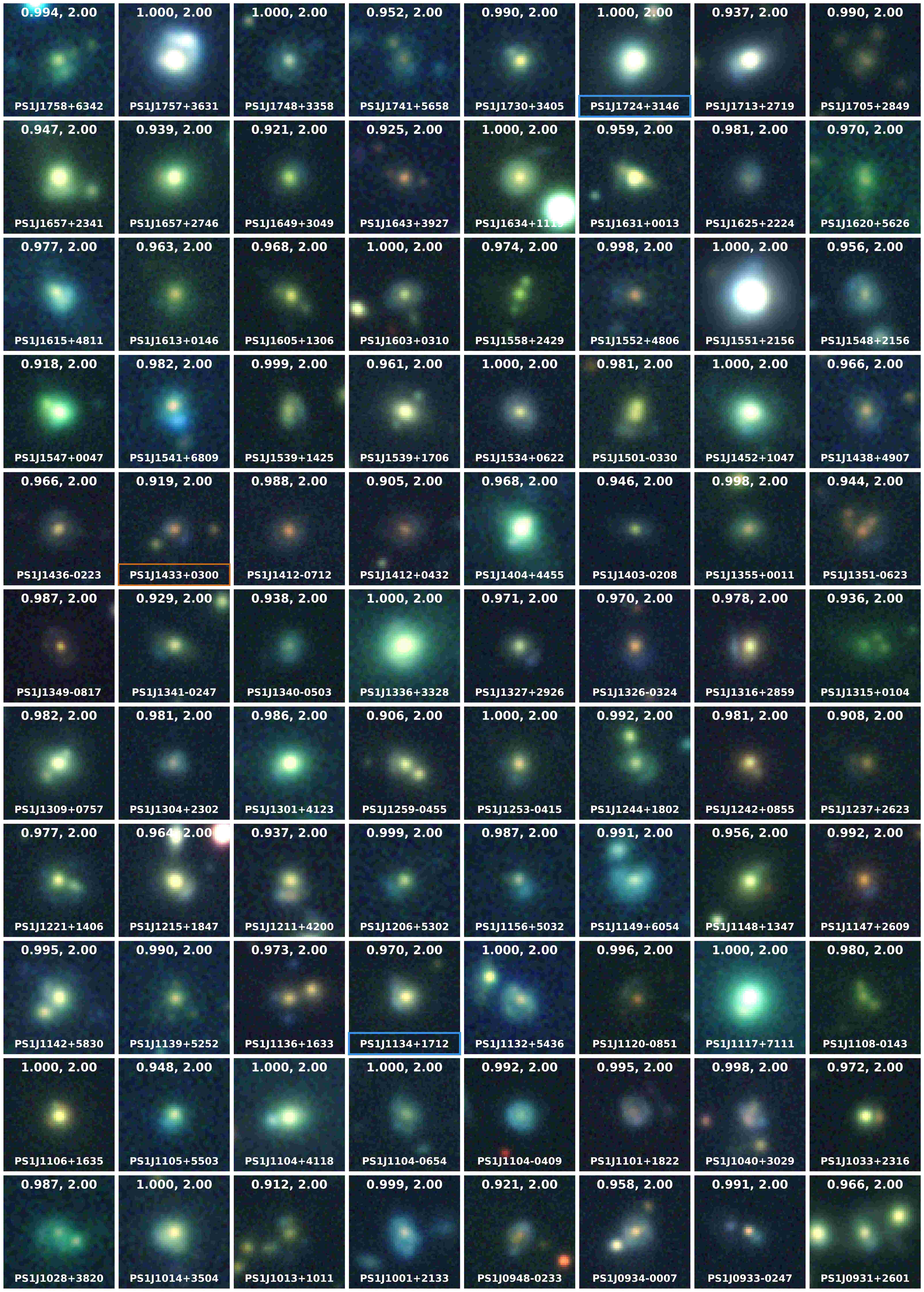}
\caption{continued.}
\end{figure*}

\begin{figure*}
\centering
\includegraphics[width=.965\textwidth]{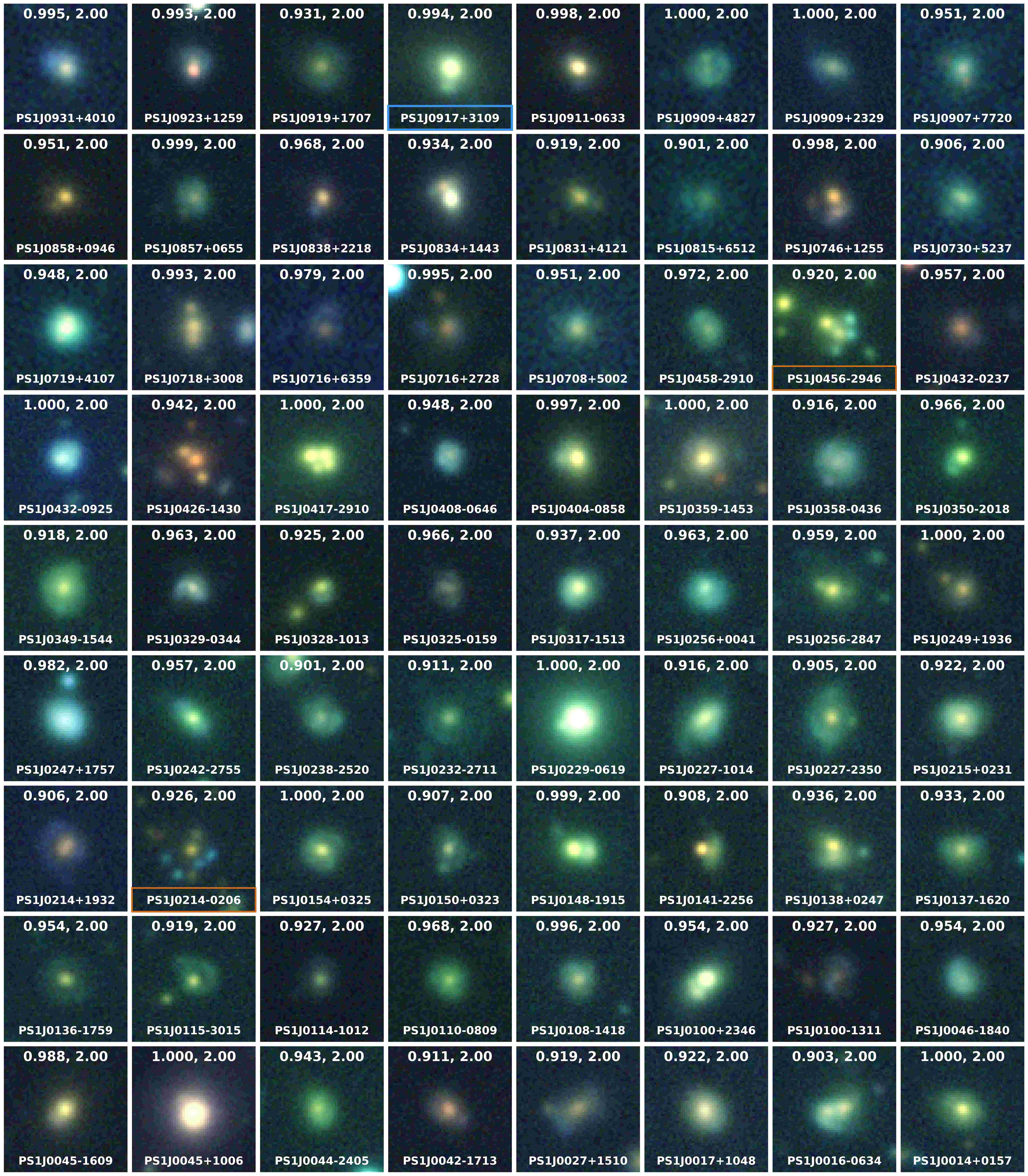}
\caption{continued.}
\end{figure*}

\begin{figure*}
\centering
  
\includegraphics[height=.28\textheight]{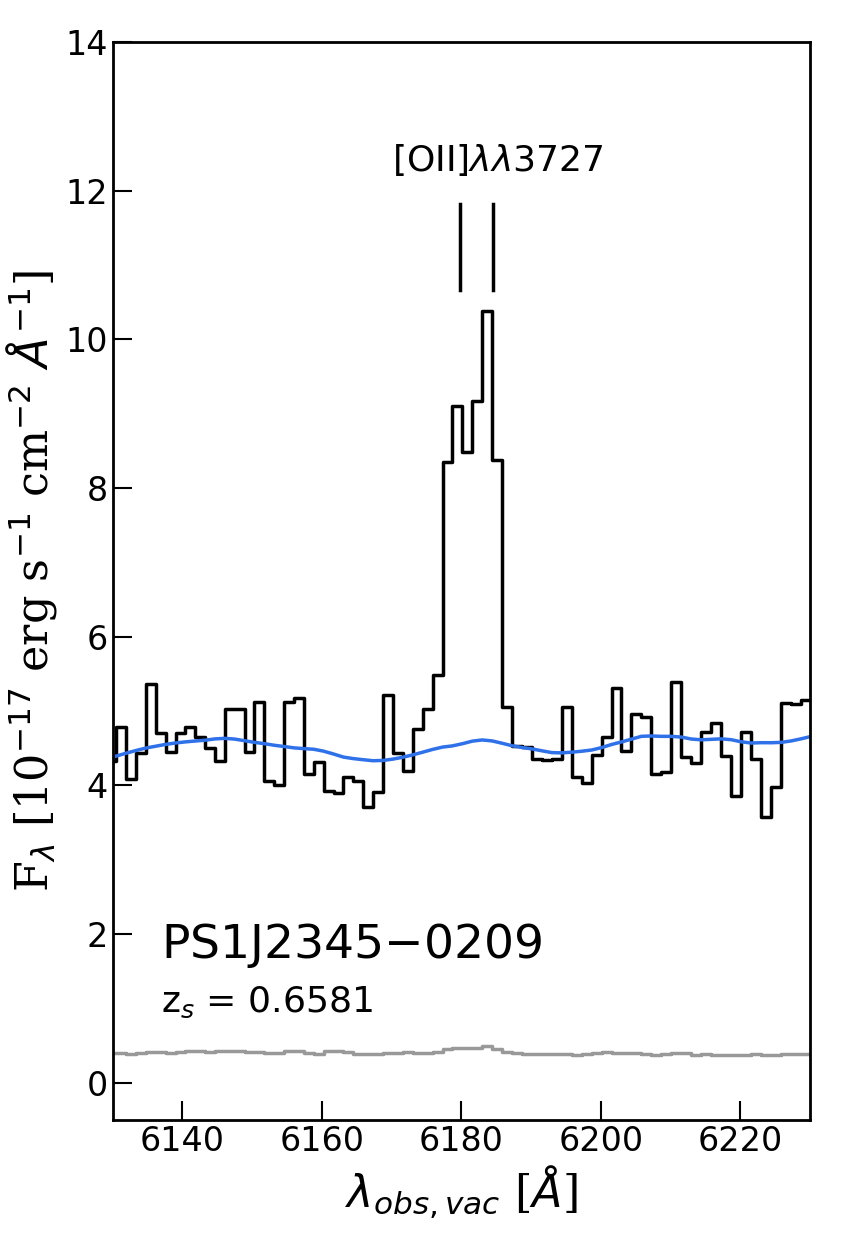}
\includegraphics[height=.28\textheight]{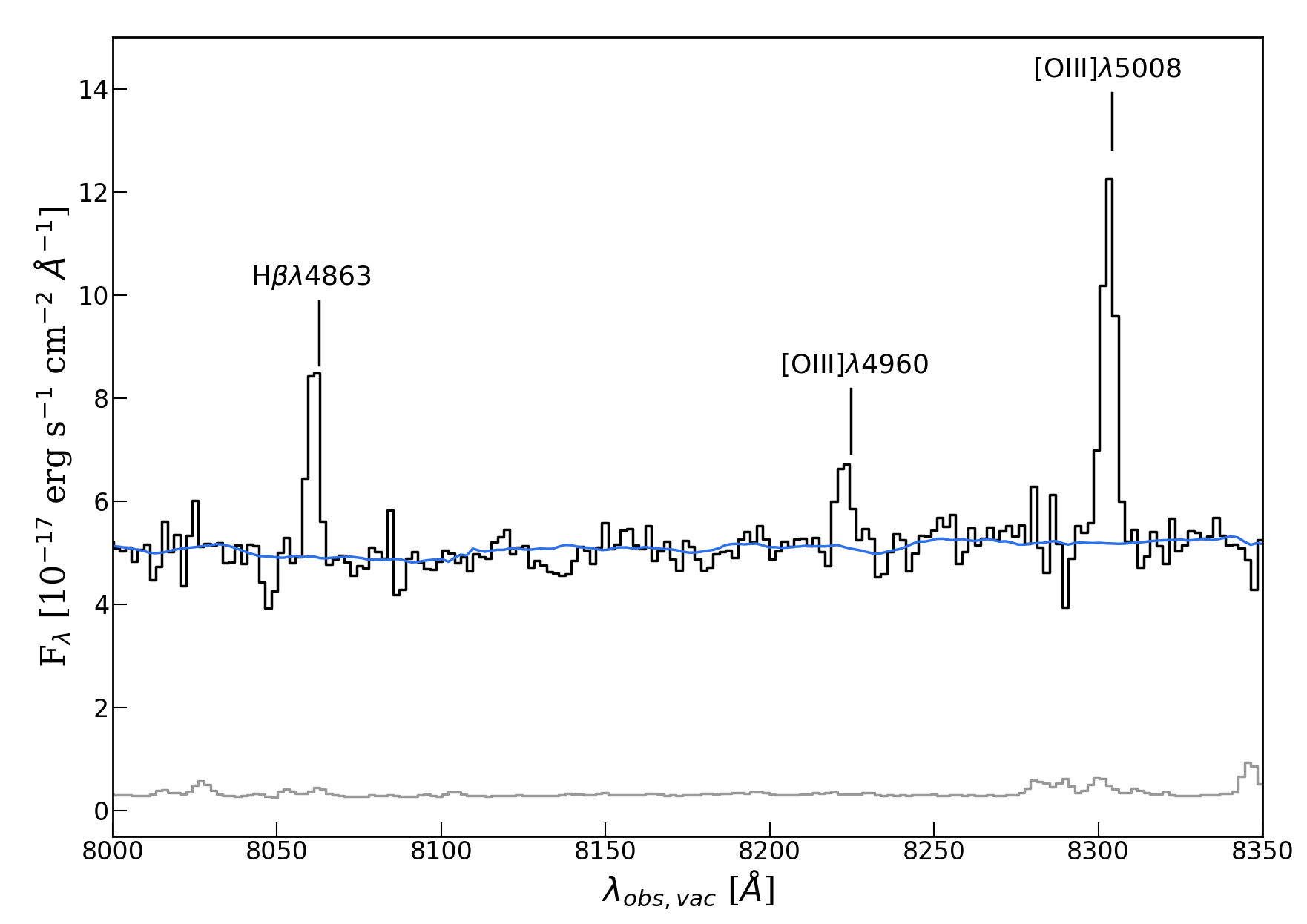}

\includegraphics[height=.28\textheight]{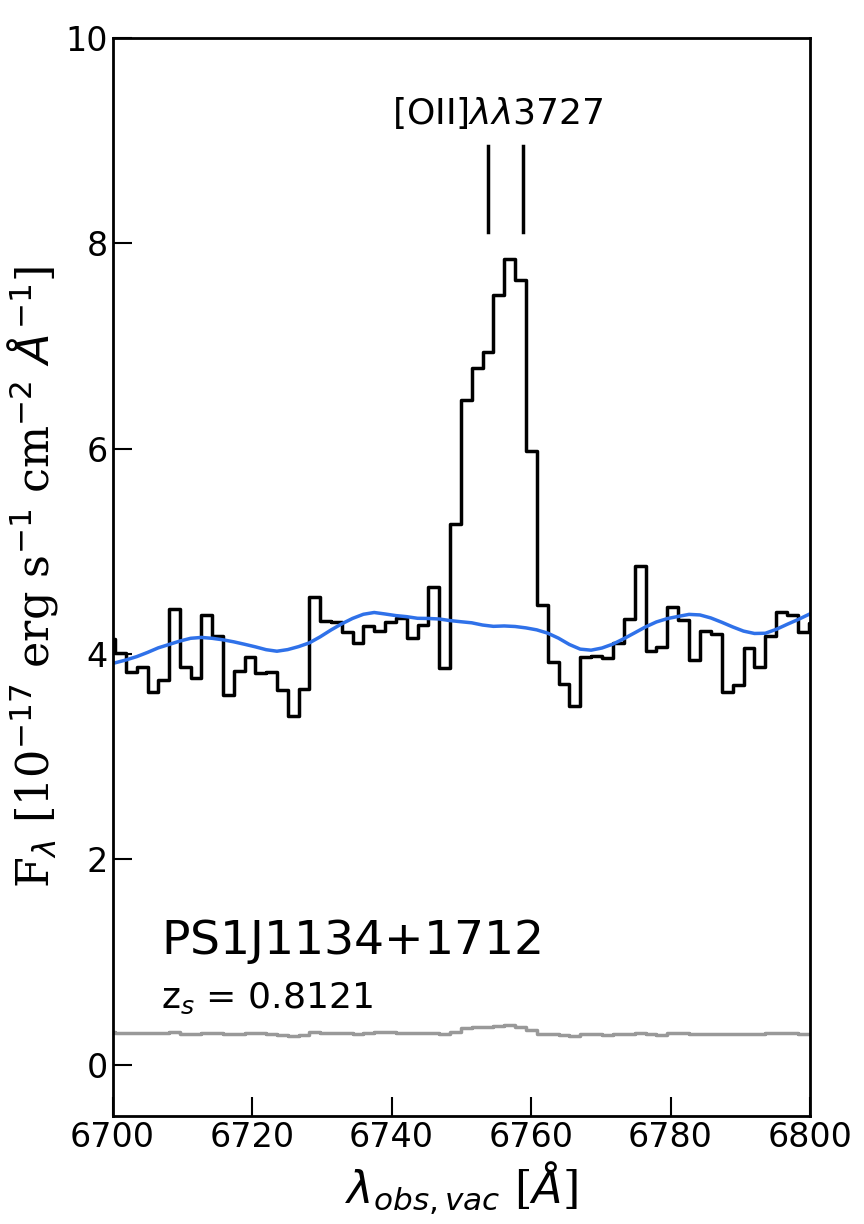}
\includegraphics[height=.28\textheight]{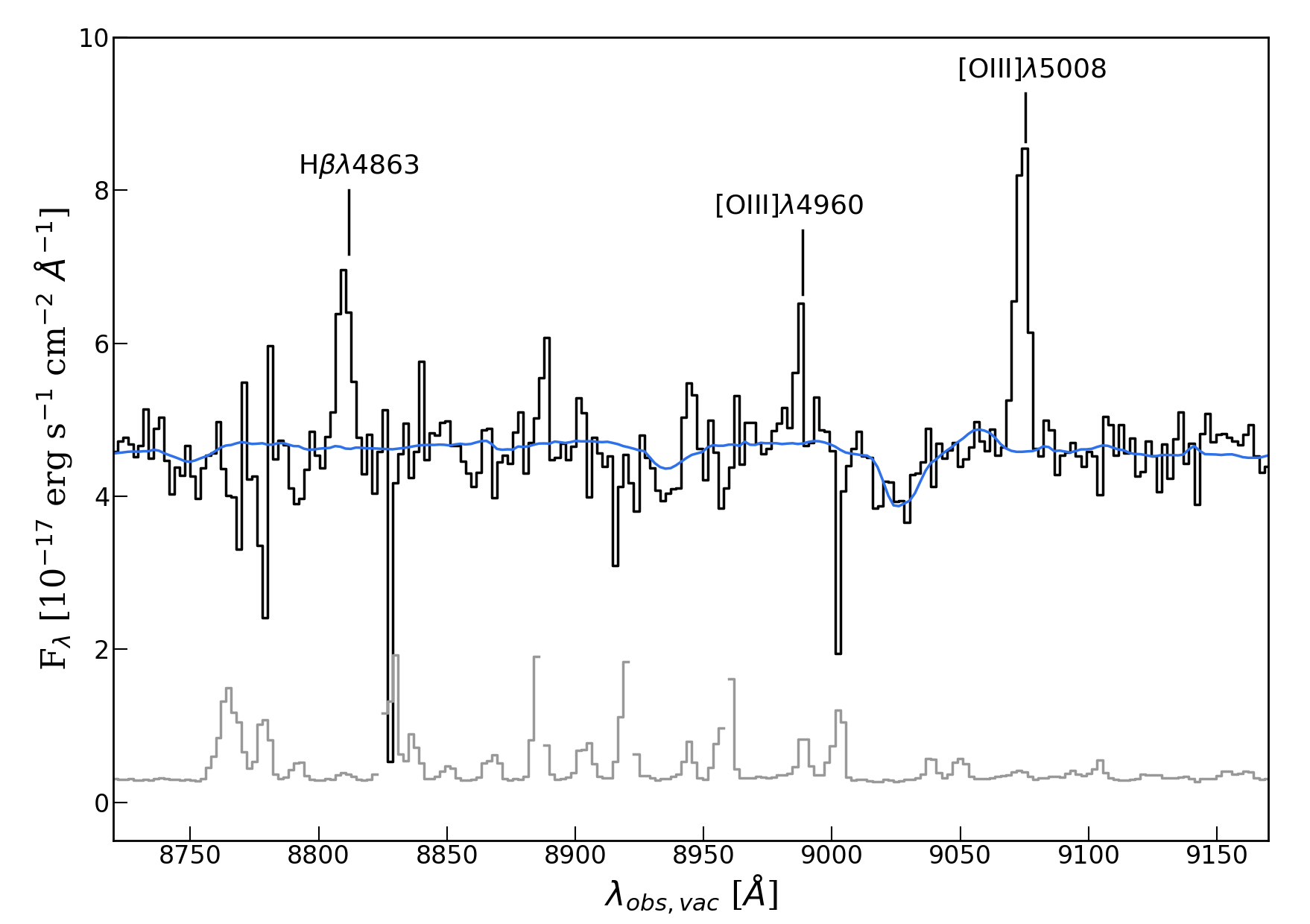}

\includegraphics[height=.28\textheight]{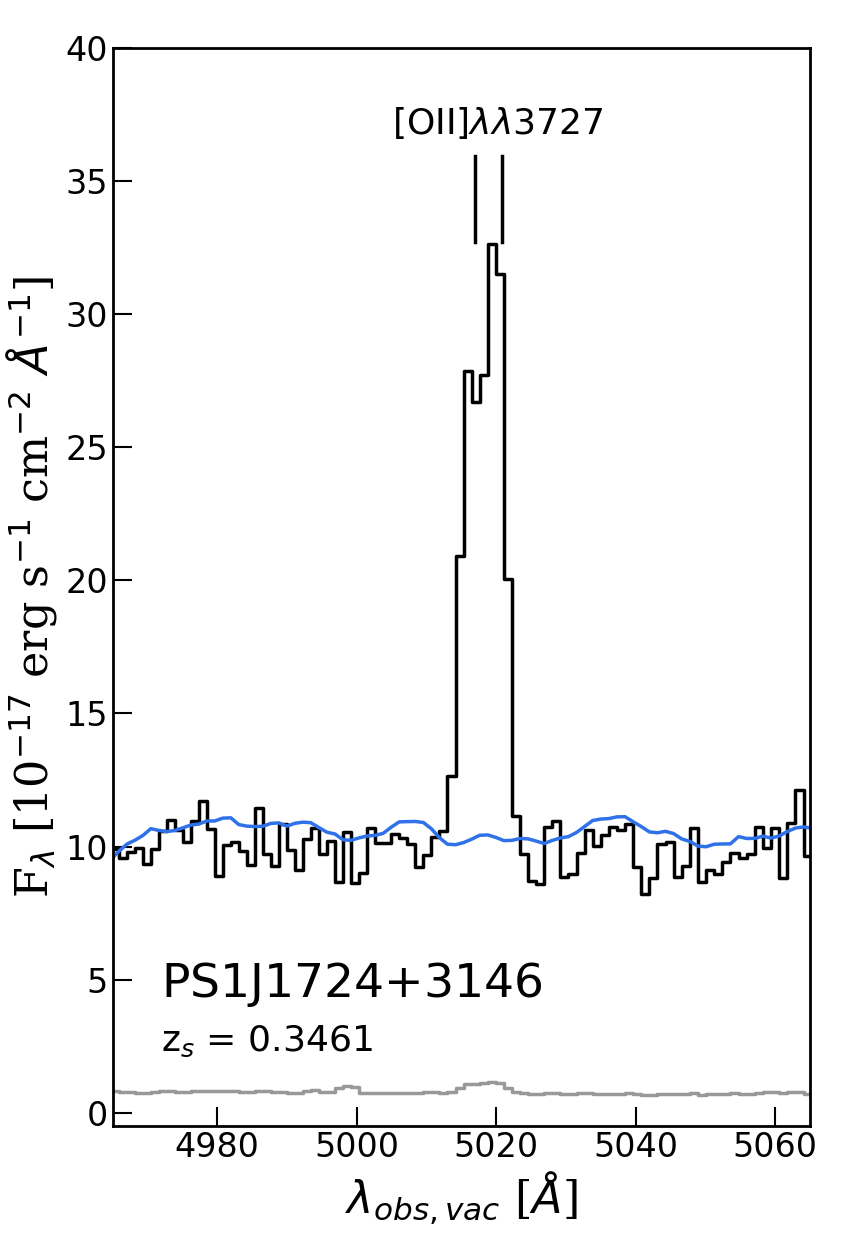}
\includegraphics[height=.28\textheight]{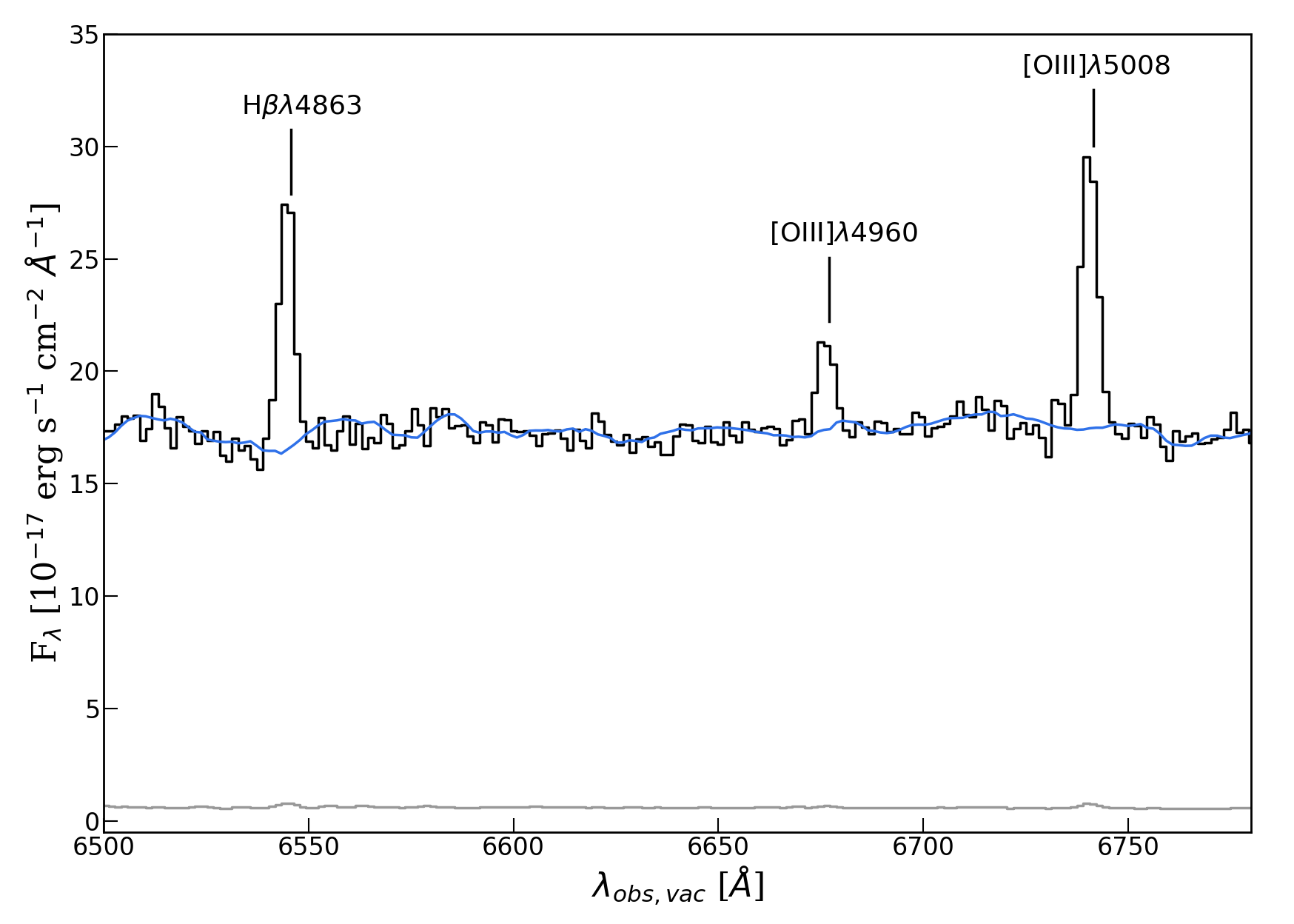}
\caption{
  PS1 lens candidates with multiple emission lines from a high-redshift background galaxy overlaid on the SDSS BOSS spectrum
  of the foreground LRG. The black, gray and blue lines show the observed spectrum, the 1$\sigma$ noise level, and the best-fit SDSS
  template for the LRG, respectively. The spectrum is zoomed on the spectral features associated with the background line-emitter
  rather than with the LRG. {\it Top:} PS1J2345$-$0209 with $z_{\rm LRG}=0.2940$ and $z_{\rm s}=0.6581$. {\it Middle:} PS1J1134$+$1712
  with $z_{\rm LRG}=0.3752$ and $z_{\rm s}=0.8121$. {\it Bottom:} PS1J1724$+$3146 with $z_{\rm LRG}=0.2097$ and $z_{\rm s}=0.3461$.}
\label{fig:boss_conf3}
\end{figure*}

\end{appendix}

\end{document}